\documentclass{article}

\newcommand{\vctr}[1]{\ensuremath{\mathbf{ #1 }}}
\newcommand{\pb}[2]{\ensuremath{\left\{ #1 , #2 \right\} }}
\newcommand{\dr}[1]{\ensuremath{\mathrm{d} #1\,}}
\newcommand{\mc}[1]{\ensuremath{\mathcal{#1}}}
\newcommand{\dbd}[2]{\ensuremath{\frac{\dr{#1}}{\dr{#2}}}}
\newcommand{\pbp}[2]{\ensuremath{\frac{\partial #1}{\partial #2}}}
\newcommand{\vbv}[2]{\ensuremath{\frac{\delta #1}{\delta #2}}}
\newcommand{\ket}[1]{\ensuremath{\left|  #1 \right\rangle}}
\newcommand{\bra}[1]{\ensuremath{\left\langle #1 \right|}}
\newcommand{\bk}[2]{\ensuremath{\left\langle #1 | #2 \right\rangle}}
\newcommand{\proj}[2]{\ensuremath{\ket{#1} \bra{#2}}}
\newcommand{\tpk}[2]{\ensuremath{\ket{#1}\!\otimes\!\ket{#2}}}
\newcommand{\matel}[3]{\ensuremath{\bra{#1} #2 \ket{#3}}}
 
\newcommand{\op}[1]{\ensuremath{\widehat{\textsf{\ensuremath{#1}}}}}
\newcommand{\opad}[1]{\ensuremath{\op{#1}^{\dagger}}}
\newcommand{\id}{\op{\mathsf{1}}}
\newcommand{\denop}{\ensuremath{\rho}}

\newcommand{\comm}[2]{\ensuremath{\left[ #1 , #2 \right]}} 

\newcommand{\nrm}{\frac{1} {\sqrt{2} } }
\newcommand{\edf}{\ensuremath{=_{_{df}}}}
\newcommand{\real}{\ensuremath{\mathrm{Re}}\,}
\newcommand{\imag}{\ensuremath{\mathrm{Im}}\,}
\newcommand{\ddt}{\ensuremath{\frac{\dr{}}{\dr{t}}}}

\newcommand{\be}{\begin{equation}}
\newcommand{\ee}{\end{equation}}

\newcommand{\ie}{\mbox{i.\,e.\,\ }}
\newcommand{\iec}{\mbox{i.\,e.\,}}

\newcommand{\eg}{\mbox{e.\,g.\,\ }}
\newcommand{\egc}{\mbox{e.\,g.\,}}
\newcommand{\etc}{etc.\,\ }

\usepackage{chicago}
\usepackage{supertabular}
\usepackage[dvips]{graphics,pstcol}
\usepackage{pst-node,pst-coil}

\makeatletter
\renewcommand{\section}{\@startsection
   {section}%
   {1}%
   {0mm}%
   {\baselineskip}%
   {0.5\baselineskip}%
   {\bfseries\normalsize\centering}}%

\renewcommand{\subsection}{\@startsection
   {subsection}%
   {2}%
   {0mm}%
   {\baselineskip}%
   {\baselineskip}%
   {\normalfont\normalsize\itshape}}%

\makeatother

\begin{document}

\begin{center}
\LARGE
\textbf{Emergence of particles from bosonic quantum field theory}

\vspace{0.3cm}

\textbf{\textit{David Wallace$^*$}}

\begin{figure*}[b]
(\textit{December 23, 2001})

* Centre for Quantum Computation, The Clarendon Laboratory, 
University of Oxford, Parks Road, Oxford OX1 3PU, U.K.

(\textit{e-mail:} david.wallace@merton.ox.ac.uk).

\end{figure*}

\normalsize
\end{center}

\vspace{0.7cm}

\begin{quote}
An examination is made of the way in which particles emerge from linear, bosonic, massive
quantum field theories.  Two different constructions of the one-particle subspace of such theories
are given, both illustrating the importance of the interplay between the quantum-mechanical linear 
structure and the classical one.  Some comments are made on the 
Newton-Wigner representation of one-particle states, and on the
relationship between the approach of this paper and those of Segal, and
of Haag and Ruelle.

\end{quote}

\vspace{0.4cm}

\newcommand{\ob}[1]{\ensuremath{\overline{#1}}}
\newcommand{\lsqr}{\mathrm{L}^2}
\newcommand{\lgr}{\ensuremath{\mathrm{L}}}
\newcommand{\ham}{\ensuremath{\mathrm{H}}}
\renewcommand{\Re}{\mathrm{Re}}
\renewcommand{\Im}{\mathrm{Im}}

\newcommand{\rfb}[1]{\psframebox{\begin{tabular}{c} #1 \end{tabular}}}

\section{Introduction}

For better or for worse, most quantum systems are found by starting with a classical 
system and then quantizing it.  The states of the resulting quantum system will be 
described by complex functions on the configuration space of the classical system, 
whose squared moduli tell us the probability density for finding the system in a given 
configuration.  

Applying this method to classical fields would seem, if not unproblematic, then at 
least difficult only for technical reasons.  We would naturally expect to find a theory 
whose states are wave-functionals on the configuration space: that is, maps which associate 
a complex number to each configuration of the classical field on a given hypersurface.

Scarcely a vestige of this behaviour is seen in the usual phenomenology of `quantum field 
theory'.  Instead we find ourselves with a theory usually described in terms of particles: 
quarks, gluons, electrons\ldots, and the localized interactions between them.  The 
field-configuration viewpoint is occasionally seen (notably in the path-integral formalism 
of quantum field theory) but is usually regarded as at best a calculational tool.

Furthermore, these particles are notoriously strange entities.  Various results of 
quantum field theory seem strongly to imply that they cannot be localized in any meaningful, 
covariant way; that they must be created and annihilated in interactions which cannot be 
spatio-temporally localized; that we cannot start with a theory of free particles and 
`turn on' an interaction without pathology (implying that the concept of particle is 
bound up with the dynamics of the theory and is not just a kinematic
concept); and that the particles which should be associated with a given
field theory vary according to the energy levels at which that theory is
studied.

For these reasons, it is normal in modern quantum field theory to regard
the field as the primary concept and the particles as secondary,
derivative entities.  This process has been studied extensively using
the methods of algebraic quantum field theory
and the signs are encouraging that it can be understood in a
mathematically and conceptually rigorous way; however, the very
abstractness of these methods can make it difficult to understand quite
why the idea of `particle' should be so powerful in understanding the
prima facie very different concepts inherent in a quantised
\emph{field}.

The purpose of this paper, then, is to analyse in a fairly concrete context the 
way in which certain subspaces of a quantum field theory's Hilbert space come to possess
characteristics of a one-quantum-particle Hilbert space.  `The concrete context' in question
is that of a massive, scalar, bosonic field, assumed to be asymptotically treatable as a free field.
Section \ref{fields} presents the classical and quantum theory of such a field, and section
\ref{defining} considers what the correct definition should be of
`local' and `particle' states in QFT.   In sections 
\ref{modal}--\ref{operator} --- which form the core of the paper --- two separate
constructions of the one-particle subspace are given, both of which
illustrate the central role played by the interaction between the linear
structure on the QFT Hilbert space (present in any quantum system) and
the linear structure on the classical phase space (specific to a linear
field theory).  Section \ref{newtonwigner} discusses the Newton-Wigner
representation of position for one-particle systems, and section
\ref{comparisons} makes some brief comments on the relationship between
this paper's approach to QFT and some other approaches; section
\ref{conclusion} is the conclusion.

Currently, it is increasingly common for foundational discussions of QFT
to be conducted in the powerful and abstract language of algebraic
quantum field theory.  This paper eschews that tendency: some concepts
and results of algebraic QFT are referred to, but the framework used here is
much closer to that used by the mainstream physics community.  For a
defence of the validity of using this apparently rather non-rigorous
framework for a foundational discussion, see \citeN{wallaceqft}.

\section{Field quantization}\label{fields}

In this section, we shall review the method by which free, and weakly
interacting, field theories are quantized.  We shall outline the
problems which occur when we try to reinterpret these quantized theories
as fundamentally about particles, and then consider, in qualitative
terms, how particles can enter the theory in a non-fundamental way.

\subsection{Classical free fields}

A fairly general second-order field equation for a free-field theory is
\be\label{fieldeqn}
\frac{\mathrm{d}^2}{\mathrm{d}t^2}\phi(\vctr{x},t) + \mc{R} \phi(\vctr{x},t) = 0
\ee
where $\phi(\vctr{x},t)$ is a real field defined on some manifold $\Sigma\times \mathrm{R}$, 
points on
$\Sigma$ are labelled by $\vctr{x}$, and 
$\mc{R}$ is some real symmetric operator acting on functions of the spatial coordinate 
$\vctr{x}$.  

The simplest example of such a theory is the real Klein-Gordon equation, for which
\be \mc{R} = m^2 - \nabla^2,\ee
and in fact if we
want a theory with relativistic covariance then there are no other
examples.  However, more general theories of this form are potential
models for:
\begin{itemize}
\item Curved spacetimes with a time translation symmetry;
\item Systems interacting with a time-independent background field;
\item Solid-state systems.
\end{itemize}
Most crucially for our purposes, field theories with this form also occur as approximations 
to nonlinear theories; we will consider this case in more detail later.

Such a theory can be generated from a Lagrangian 
\be \lgr[\phi,\dot{\phi}] = \frac{1}{2}\int_{\Sigma} \dr{^3\vctr{x}} \left( \dot{\phi}(\vctr{x})^2
- \phi(\vctr{x})(\mc{R}\phi)(\vctr{x})\right).\ee 

Carrying out the Legendre transform to the Hamiltonian formalism, we get a set of canonical coordinates
$\phi(\vctr{x})$ labelled by $\vctr{x}$, a set of conjugate momenta
$\pi(\vctr{x})$, and a Hamiltonian \ham, where 
\be \pi(\vctr{x}) \edf \vbv{\lgr}{\dot{\phi}(\vctr{x})} = \dot{\phi};\ee
\be \label{hamilt}\ham [\phi,\pi] = \frac{1}{2}\int_{\Sigma} \dr{^3\vctr{x}} \left(
\pi(\vctr{x})^2 + \phi(\vctr{x})(\mc{R}\phi)(\vctr{x})\right).\ee 
Points in the phase space \mc{P} of the field are then given by
specifying pairs of functions $(\phi,\pi)$; the Poisson bracket on \mc{P} is given by
\be \pb{A[\phi,\pi]}{B[\phi,\pi]} = 
\int_\Sigma\dr{^3\vctr{x}}\left( \vbv{A}{\phi(\vctr{x})}\vbv{B}{\pi(\vctr{x})} -
\vbv{A}{\pi(\vctr{x})}\vbv{B}{\phi(\vctr{x})}\right),\ee
so of course $\phi$ and $\pi$ obey the canonical relations
$\pb{\phi(\vctr{x})}{\phi(\vctr{y})}=\pb{\pi(\vctr{x})}{\pi(\vctr{y})}=0$
and $\pb{\phi(\vctr{x})}{\pi(\vctr{y})}=\delta(\vctr{x}-\vctr{y}).$

Through each point in phase space flows a unique trajectory; hence
points in \mc{P} are in one-to-one correspondence with solutions of
(\ref{fieldeqn}).

\subsection{Field quantization}\label{fieldquantization}

We will quantize classical fields (free or interacting) in the most naive possible way: by 
direct comparison with non-relativistic particle mechanics.  That is, we
will represent states of the quantum system by complex wave-functions
on the configuration space of the classical system.  In this case, that
configuration space is the infinite-dimensional space \mc{S}
of functions on $\Sigma$, 
so the quantum states will be functionals $\Psi[\chi]$ on this space 
(we will denote the Hilbert space of all such functionals as
$\mc{H}_\Sigma$). By
analogy with the non-relativistic quantization of the coordinates $q,p$ as  
\be  \op{q} \psi =q \psi(q); \,\,\,\op{p} \psi = -i \dbd{\psi}{q}\ee
we will quantize the coordinates $\phi(\vctr{x})$ and $\pi(\vctr{x})$ as
\be (\op{\phi(\vctr{x})} \Psi)[\chi] = \chi(\vctr{x})\Psi[\chi];\ee
\be (\op{\pi(\vctr{x})} \Psi)[\chi] = - i
\vbv{\Psi}{\chi(\vctr{x})}[\chi].\ee  It is easy to check that the
canonical commutation relations are satisfied:
\be \comm{\op{\phi(\vctr{x})}}{\op{\phi(\vctr{y})}} = 
\comm{\op{\pi(\vctr{x})}}{\op{\pi(\vctr{y})}}=0;\ee
\be \comm{\op{\phi(\vctr{x})}}{\op{\phi(\vctr{y})}} = i
\delta(\vctr{x}-\vctr{y}).\ee

It is to be admitted that we have been very cavalier with our treatment
of the infinite-dimensional spaces in use here.  It is possible (whilst
we confine ourselves to free fields) to be much more careful and
rigorous,\footnote{See \citeN{marsden}; \citeN{woodhouse} for discussions of infinite-dimensional
classical mechanics, and \citeN{waldqft} for a careful discussion of quantising linear field theories.}
but if we wish our framework to be powerful enough to handle interactions 
then there is actually no need for infinite-dimensional technicalities,
for reasons to be explained in section \ref{renormalisation}.

\subsection{Interactions and renormalisation}\label{renormalisation}

Formally speaking, nothing in the previous description will be altered
if we add some higher-order terms (such as $\phi^4$), which change the
field equation from free to interacting: we could restrict our attention
to regimes in which these terms are small in comparison to the free-field 
Hamiltonian, and proceed to analyse their effects using
perturbation theory.

However, the reader may at this stage object that we are playing fast
and loose with some very poorly-defined mathematical concepts.  In fact,
it is well-known that terms like $\phi^4$, when added to the
Hamiltonian, give contributions which are not small, but infinite ---
hence formulating a well-defined interacting quantum theory is actually
very subtle.  In fact, one approach would be to say that the only
quantum theories we understand well enough for conceptual study are the
free-field ones, and confine our attention to those.

In this paper, however, we shall take a more liberal attitude.  There is
actually a well-defined approach to understanding these apparent
infinities, worked out primarily by Kenneth Wilson and originating in
solid-state physics.   In Wilson's approach, we postulate that QFTs do not after all
have infinitely many degrees of freedom; rather, some unknown processes
cut off the high-energy degrees of freedom and leave only finitely many
to contribute to the physics.  It then turns out --- rather remarkably 
--- that all interaction terms in the Hamiltonian will fall into two categories.  
\emph{Non-renormalisable} interactions will be negligibly weak on energy scales
far lower than the cutoff threshold.  \emph{Renormalisable} interactions
are not necessarily negligible, but at low energies they are affected by
the choice of the cutoff only through modifications (``renormalisation'') of the
parameters in the interaction terms.  Since these parameters are in any
case only known through experiment, the choice of the cutoff becomes
irrelevant to the low-energy regimes of the QFT.

Solid-state physics provides an example of this process.  If we study a
solid-state system on length-scales which are large compared to the
interatomic spacing, we can approximate the possible (classical) configurations
of the atoms by a continuous function --- and thus approximate the
system by a continuous field theory.  In quantizing this theory we find
that interaction terms lead to infinities, but these are an artefact of
our continuum assumption.  Once we introduce a cutoff banning
excitations of the system which vary significantly on length-scales
short in comparison with the interatomic separation, the infinities
vanish.  

Because we are understanding field theories in this way, we can take a
relaxed attitude to the infinite-dimensional spaces which we will
encounter in our analysis: such spaces are `really' finite-dimensional,
with the very short-distance excitations disallowed.  As for the
interaction terms, we will not have need of their specific forms.
We shall just assume, where necessary, that such terms are present but
that the theory has been renormalised and that, after renormalisation,
the interaction terms can be treated perturbatively.  For details of the
mathematics of this process, see \citeN{peskinschroeder} or any other QFT
textbook; for a conceptual discussion see \citeN{wallaceqft}.

\subsection{Problems with a particle interpretation}\label{problems}

The theory constructed above is undeniably a field theory, in the sense
that its configuration space, and fundamental observables, are
inherently field-theoretic.  It is, however, tempting to try to
reinterpret the theory so as to make direct contact with the particle
concept, either by establishing some kind of `duality' between field and particle
 descriptions (in the same sense that there is a duality between
position and momentum representations in ordinary quantum mechanics,
with neither representation being privileged over the other) or by
replacing the field description entirely with a particulate one (in
which case, presumably, the field observables would just count as
auxiliary constructions of no direct physical significance).  

There are however, many problems which emerge as soon as we try to
interpret any QFT so as to incorporate particles at a fundamental level:
\begin{itemize}
\item The `elementary particles' of particle physics are generally
understood as pointlike objects, which would seem to imply the existence
of position operators for such particles.  However, if we add the
requirement that such operators are covariant (so that, for instance, a
particle localised at the origin in one Lorentz frame remains so
localised in another), or the requirement that the wave-functions of the
particles do not spread out faster than light, then it can be shown that
no such position operators exist.  (See \citeN{halvorsonclifton}, and
references therein, for details.)
\item In non-relativistic quantum mechanics, it is straightforward to
construct Hamiltonians which describe particles interacting via 
long-range forces (for a simple example, consider two charged particles interacting 
via a Coulomb force).
However, the concept of a long-range interaction \emph{prima facie} requires some sort 
of preferred reference frame, which seems to cast doubt upon the possibility of 
constructing such
an interaction in a relativistically covariant way.  
\item As was mentioned in section \ref{renormalisation}, if interactions
are present in a QFT then it is necessary to work, not with the bare
parameters in the Hamiltonian, but with `renormalised' parameters ---
and the parameters which must be renormalised include some of those,
such as charge, which are generally taken to be intrinsic properties of
particles.  However, there is no privileged way of renormalising the
parameters, so that the values of these parameters --- and hence, the natures of the particles
which they purport to describe --- can be in part a purely
conventional matter.  
\item When we consider quantum field theory on a general spacetime
background, there is no unique procedure to define particles, and states
which appear particulate in one reference frame do not do so in other
reference frames.  For instance, consider the so-called `Unruh effect',
in which the Minkowski vacuum of a free QFT looks like a thermal (hence,
non-particulate) state to a uniformly accelerating observer.  In this
example it may be possible to argue that non-inertial observers'
descriptions are somehow less fundamental, but in a less symmetric
spacetime there will be no preferred class of observers available, hence
no preferred definition of particle.  (For a more detailed account of
this point, see \citeN{waldqft}, who advocates abandoning the particle
concept as a consequence.)
\end{itemize}

Not all have abandoned particles as fundamental in view of these
difficulties: Fleming has given a strong defence of the idea that
particle localisation does indeed make sense in relativistic QFT (see \citeN{fleming}, 
\citeN{flemingbutterfield}, and references therein) and
Weinberg's recent QFT textbook \cite{weinberg} explicitly begins with particles
and constructs the fields as auxiliary objects.  However, the general
consensus in QFT (insofar as such issues are ever explicitly addressed\footnote{See 
\cite{wilczek} for an explicit statement of this consensus.})
appears to be that the subject is primarily about quantum \emph{fields}.  In fact, much 
modern research in the field only really makes sense from this
viewpoint: for example, consider lattice quantum chromodynamics (which
attempts to understand quark confinement and the existence of protons
and neutrons, but is formulated in terms of field configurations and
makes only limited contact with the elementary heuristic that a proton is `just' three
particulate quarks bound together); or consider the quantum Sine-Gordon
equation \cite{coleman}, which has two distinct particle descriptions
(one fermionic, one bosonic) with the weak-field version of the one
equivalent to the strong-field version of the other).

Of course, none of this is to deny that particles \emph{exist}, merely
that they are not part of the fundamental ontology of quantum field
theory.  In the next section we will consider how it might be possible for the particle concept
to be recovered from a field-theoretic description.

\subsection{The particle as emergent concept}

It is a central result of condensed-matter physics that, if we start with some 
macroscopic collection 
of nonrelativistic particles close to some collective stable state, 
small excitations from that state can often be treated in terms of creating 
`particles'.  It is also generally true that, for strongly interacting systems, 
these `particles' do not coincide with the particles from which the system is 
built: so vibrations in a crystal are described in terms of `phonons', which 
are not crystal atoms, and quantized waves in a magnet are described in terms of 
`magnons' which are not iron atoms \cite{kittel2}.

There are striking formal parallels with quantum field theory: in fact, the 
construction of phonons from a monatomic crystal is virtually the same as the 
construction of particle states in a massless, scalar quantum field theory.  The 
difference is, the ontology of a crystal is not in question.  It is definitely 
made up of the lattice atoms - which correspond to the field states at different 
space points in scalar QFT.  Nonetheless many phenomena can be described by regarding 
the crystal as a gas of phonons, and some --- \eg heat transport --- require us to 
think in terms of localized phonons \cite{kittel1}.  

There is nothing particularly paradoxical about this: the crystal isn't
`really' a gas of phonons, it's just that certain states of the crystal
have properties very similar to such a gas, and that treating these
states as such is a great boon to analysis of crystal dynamics.  This
puts phonons and their ilk in good company, for a great many objects in
science --- such as animals, or rigid bodies --- have to be understood
in the same way.  There are no perfectly rigid bodies, for instance (and
they are certainly not part of the basic ontology of any fundamental
physical theory), yet certain states of a many-particle system
approximate the behaviour of `ideal' rigid bodies extremely well, and so
deserve the name.  (See \citeN{wallacestructure} for a more detailed
discussion of this point.)

We shall adopt the same attitude to the particles of relativistic quantum 
field theory: that is, we shall 
look for subspaces of the QFT Hilbert space in which the states have
particulate properties.  This will require us to formulate a definition
of `particle' and then to show that there are states of the QFT which
approximately satisfy that definition; the rest of the paper is
concerned with this task.  First, though, we need to consider in which
situations we would expect a QFT to appear particulate.

\subsection{Particle regimes}

The phenomenology of quantum field theory suggests two regimes in which
we expect particle behaviour:
\begin{itemize}
\item \textbf{The non-relativistic limit}, in which the QFT appears to
be described by slow-moving particles interacting by long-range forces;
\item \textbf{The scattering limit}, in which particles begin widely
separated, interact by short-range forces, and at late times are again
found in widely separated states.
\end{itemize}
We shall be concerned almost exclusively with the second case, for
reasons of mathematical tractability rather than on conceptual grounds:
the analysis of relativistic fields via the methods of scattering theory
is fairly well understood, whereas the process by which nonrelativistic
quantum mechanics emerges as a limiting case of QFT is much more
complicated.  In the case of scattering theory, though, at times sufficiently
long after (or before) the scattering event, the theory becomes very
well approximated by a free quantum field theory.  (This is intuitively
plausible since for scattering theory to be applicable in the first place 
it is necessary that the nonlinear terms in the Hamiltonian constitute,
after renormalisation, only a small perturbation to the free-field
theory; for a much more careful discussion and justification, see
\citeNP{haag}.)

For this reason, our analysis henceforth will be restricted to free
quantum theories (more specifically, to quantum theories of the form
(\ref{fieldeqn}); this includes some sorts of background-field
interactions).

\section{Defining particles}\label{defining}

In this section, we shall work out a definition of what properties a
family of QFT states ought to have in order to count as `particle'
states.  Since the idea of `particle' is plainly at least connected to
the concept of a localised state, we begin by considering how the latter 
states are to be defined in QFT.

\subsection{Localised states in a field ontology}\label{local}

Which field-theory states are to count as localised?  

In a QFT the idea of localisation must enter through the spatial localisation of the 
observables.
The observables of the theory are defined via the field operators $\op{\phi}(\vctr{x},t)$ and 
$\op{\pi}(\vctr{x},t)$, so it is natural to define any given observable at time
$t$ as being localised in a spatial region $\Sigma_i\subseteq \Sigma$ iff it is a function
only of field operators of form $\op{\phi}(\vctr{x}_i,t)$ and
$\op{\pi}(\vctr{x}_i,t)$ with all of the $\vctr{x}_i$ in $\Sigma_i$.

But if defining localised observables is straightforward, defining
localised states will prove decidedly less so.  We might begin by
trying:

\begin{quote}\textbf{Naive localisation}: A state \ket{\psi} is localised in a
spatial region $\Sigma_i$ iff $\matel{\phi}{\op{O}}{\phi}=0$ for any observable
$\op{O}$ localised outside $\Sigma_i$.
\end{quote}

This seems plausible when we compare it with the classical case: there a
state is localised in $\Sigma_i$ if $\pi(\vctr{x})=\phi(\vctr{x})=0$ for any
$\vctr{x}\notin \Sigma_i$.  But it is mathematically impossible for any state 
to satisfy it, for it
implies that for any such $\vctr{x}$, and for any $n \in \mc{Z}^+$, 
\be
\matel{\psi}{\op{\phi}^n(\vctr{x},t)}{\psi}=\matel{\psi}{\op{\pi}^n(\vctr{x},t)}{\psi}=0.\ee
But this would imply that \ket{\psi} was a simultaneous eigenstate of
$\op{\pi}(\vctr{x})$ and $\op{\phi}(\vctr{x})$, and these operators have
no eigenstates in common. (The mathematics, bar some need to regularise
to deal with operators defined at a point, is the same as for the
nonrelativistic operators $\op{X}, \op{P}$, which are well-known to have
no eigenstates in common.)

Physically it is easy to see what is happening here.  The vacuum state of a
field theory (which we will denote by \ket{\Omega}) is not `nothingness', or 
`empty space'; it is simply a
slightly colourful way of describing the ground state of the field's
Hamiltonian.  In solid-state systems
(which, recall, we are treating as field-theoretic systems like any other)  
this state is just the zero-temperature state of the solid, in which the
atoms will not be at rest but will have zero-temperature fluctuations;
the same will be true for the field excitations of a relativistic field
theory.  

This suggests, however, an alternative definition, first proposed (for spacetime regions
\mc{O}, not spatial regions $\Sigma_i$) by
\citeN{knight}:

\begin{quote}\textbf{Knight localisation:} a state is localised in
a spacetime region \mc{O} iff 
$\matel{\phi}{\op{A}}{\phi}- \matel{\Omega}{\op{A}}{\Omega}=0$ for any observable
$\op{A}$ localised outside the light cone of \mc{O}.
\end{quote}

It is possible to find states satisfying this criterion \cite{knight}:
take any unitary operator \op{U} localised in \mc{O}, then the state
$\op{U}\ket{\Omega}$ will be Knight-localised in \mc{O}. 

However, Knight localisation differs in one important respect from the
sort of localisation which we encounter in NRQM.  In the latter,
properties like `is localised in \mc{O}' are treatable in the same way as properties
like `has energy $E$' or `has momentum less than $p$': that is, we can define a 
projection operator whose intended interpretation is `localised in
\mc{O}', whose range is the space of all such states.  This would be possible for 
Knight-localised states iff
they form a subspace: that is, iff any superposition of two states
Knight-localised in \mc{O} is also Knight-localised in \mc{O}.  

The fact that Knight-localised states do not have this property is a
consequence of the Reeh-Schlieder
theorem \cite{reehschlieder}.
\begin{quote}\textbf{Reeh-Schlieder theorem}: for any region \mc{O},
the set of all states generated by the action
of operators localised within \mc{O} upon the vacuum, spans the Hilbert
space of the QFT.
\end{quote}
(For a proof, and further discussion, see \citeNP{haag}.)
It follows\footnote{To see that it follows, we need only note that the unitary 
elements of a (bounded) operator algebra $\mc{A}(\mc{O})$ span $\mc{A}(\mc{O})$.  This can be proved as 
follows: for any bounded Hermitian element $\op{H}$ of $\mc{A}(\mc{O})$, and any $t\neq 0$, $(it)^{-1}(\exp(i
t \op{H}) - \id)$ is a linear combination of unitary elements of $\mc{A}(\mc{O})$.  As
$t\rightarrow 0$, this sequence tends to \op{H}, hence \op{H} is in the
span of the unitary operators.  To complete the proof, simply recall
that any linear operator can be written as $\op{A}+ i \op{B}$, where
\op{A} and \op{B} are Hermitian.}
from the Reeh-Schlieder theorem that states
Knight-localised at \mc{O} span the entire state space, which rules out
any possibility of a projector meaning `localised with certainty in \mc{O}'.

It is easy to see --- again by analogy to the solid state --- why these
problems occur.  For in a generic solid-state system, atoms are coupled
to their neighbours, and as a consequence the ground state of the system
is highly entangled.  This allows us (in principle) to exploit the 
long-range correlations between spatially separated subsystems of the field to produce any state
by local operations within \mc{O}.

(To see this process in a far simpler system, consider the four-
dimensional Hilbert space $\mc{H}_A \otimes \mc{H}_B$, where $\mc{H}_A$
and $\mc{H}_B$ are each one-qubit (two-state) systems.  The entangled
states
\be \ket{\phi\pm} = \nrm (\tpk{1}{1}\pm \tpk{0}{0})\ee
are totally indistinguishable from one another when restricted to either
subsystem (they both induce the reduced state
$\denop=\frac{1}{2}(\proj{0}{0}+\proj{1}{1})$ on each subsystem) but
their sum $\nrm(\ket{\phi+}+\ket{\phi-}=\tpk{1}{1}$ is clearly
distinguishable from both of them on either subsystem.  Examples of
this kind are analysed in rather more detail by \citeN{redhead} and \citeN{clifton}.)

However, \emph{in practice} the correlations due to vacuum
entanglement usually drop off fast enough that
using Knight-localised states to approximate states localised far from
\mc{O} requires prohibitively high-energy states.  We can then use the
following pragmatic criteria to characterise locality:

\begin{quote}\textbf{1. Effective localisation (qualitative form):} A state \ket{\psi} is 
\emph{effectively localised} 
in a spatial region $\Sigma_i$ iff for any function $\op{f}$ of field operators
$\op{\phi}, \op{\pi}$,
$\matel{\psi}{\op{f}}{\psi}- \matel{\Omega}{\op{f}}{\Omega}$
is negligibly small when $\op{f}$ is evaluated for field operators outside $\Sigma_i$, 
compared to its values when evaluated for field operators within $\Sigma_i$.
\end{quote}

\begin{quote}\textbf{2. The effective localisation principle (ELP) (qualitative form}: A
subspace \mc{H} of the QFT Hilbert space $\mc{H}_\Sigma$ obeys the ELP
on scale $L$ iff for any spatial region \mc{S} large compared with $L$, a superposition
of states effectively localised in \mc{S} is effectively localised in effectively the same 
region.
\end{quote}

These qualitative notions can be made precise in a number of ways, such as:
\begin{quote}\textbf{1. Effective Localisation (quantitative form):} A state is 
\emph{L-localised} in a region $\Sigma_i$, iff for any function $\op{f}$ of field operators
$\op{\phi}, \op{\pi}$,
$\matel{\psi}{\op{f}}{\psi}- \matel{\Omega}{\op{f}}{\Omega}$
falls off for large $d$ like (or faster than) $\exp(-d/L)$, where $d$ is the distance 
from $\Sigma_i$
at which the function $\op(f)$ is evaluated.  (Note that there is no difference, 
according to this definition, between a state $L$-localised at some spatial point 
\vctr{x} and a state $L$-localised in a region of size $\sim L$ around \vctr{x}.)
\end{quote}

\begin{quote}\textbf{2.  ELP (quantitative form):} A state obeys the ELP on scale 
$L$ iff, for any 3-sphere \mc{S} of radius $>L$, a superposition of states $L$-localised 
in \mc{S} is $L$-localised in \mc{S}.
\end{quote}

A subspace of states for which ELP holds on scale $L$ can be treated --- approximately 
--- as possessing a well-defined
concept of localisation and of ``localised in $\Sigma_i$'' projectors for regions
large compared with $L$ (these are constructed, for each such region $\Sigma_i$, by taking the 
projector onto the set
of all states in \mc{H} which are effectively localised in $\Sigma_i$; because of ELP, this set
must be a linear space).
Effectively, in such a subspace we are excluding enough states that for any sufficiently
large $\Sigma_i$, we
cannot construct states localised far from $\Sigma_i$ using only those
states localised within $\Sigma_i$.

It is still reasonable to ask: what good is effective locality?  A state
effectively localised in \mc{A} can still in principle be distinguished
from the vacuum via measurements made arbitrarily far away from \mc{A}.
This question lies rather outside the scope of this paper (see 
\citeN{halvorsonclifton} and \citeN{wallaceqft} 
for further discussion).  Here we note only that such
problems are by no means new to relativistic quantum theory.  Even in
non-relativistic quantum mechanics, there are in general no states which
remain exactly localised in a finite region for any finite period of
time --- yet this does not seem to get in the way of the concept of
localised particle in NRQM.  

For the purpose of this paper, we shall treat effective localisation as
`good enough', and (since no particularly useful concept of exact localisation exists)
will often drop the word `effective', treating effectively localised states
simply as localised.

\subsection{What is a quantum particle?}\label{particle}

Granted that a quantum field theory must be treated as being
fundamentally about fields, what properties must a given state of a
quantum field theory have in order to be deemed a particle state?  It is
instructive to start by considering the classical case: which classical
field configurations (if any) could be described as particles?  Here the
answer seems obvious: the `particle' configurations will be field
configurations which are localised in a fairly small spatial region ---
localized blobs of field, in fact.  Translated into quantum mechanics, this 
would make `particles' just
another name for the effectively localised states of the last section,
provided that they were localised to sufficiently small regions.

However, this classical concept of particle is in one sense too
weak to be appropriate for quantum theory.  
Classical wave-packets tend to spread out with time, becoming less
localised --- and hence, less `particulate', whereas in non-relativistic quantum
mechanics a state describing $n$ particles at time $t$ will continue to
describe $n$ particles at all other times --- and even in relativistic
quantum mechanics we wish to recover a notion of particle which is
robust and time-independent provided the particles are far away from one
another.\footnote{We can find classical field theories which contain states like these ---
the solitons of the sine-Gordon equation are one example \cite{coleman} --- but in
general they occur only in strongly non-linear theories, whereas here we
are concerned with linear or nearly linear theories.}

Furthermore, the criterion that particles should be
localised is in some sense also too strong for quantum mechanics.  As
the two-slit experiment reminds us, it is easy for a particle to enter a
state which is nowhere near an eigenstate of position --- in other
words, nowhere near localised.

However, the two-slit experiment also suggests the correct quantum
definition of particle.  Although the experiment shows --- by
demonstrating interference of the particle wave --- that a 
classical-particle picture  isn't viable, it also shows
that a classical-wave picture isn't viable either, because on
measurement the particle is always found to be localised \emph{somewhere}.  
To ensure within the formalism of quantum physics that this happens, it is enough
to require the particle to be a linear superposition of states all of which are localised ---
then any measurement of particle position will always give a single
answer.   (I stress that this is intended to be an essentially interpretation-independent
statement: I am not addressing the measurement problem here.)

These observations motivate our definition of a quantum
particle:\footnote{It should be noted that this definition is closely
related to the definition used in algebraic QFT, in which an $n$-particle 
state is defined as one which is able to trigger up to, but no
more than, $n$ detectors at a time.  See \citeN[section II$.4$ and chapter VI]{haag} 
for more on this
definition.}

A space of one-particle states of size $L$ (where $L$ is small), written $\mc{H}_{1P}$, is a subspace of
the QFT Hilbert space $\mc{H}_\Sigma$ such that \label{particledef}
\begin{enumerate}
\item There is a basis for $\mc{H}_{1P}$, each member of which is a state  
$L$-localised at a point; equivalently, all states in $\mc{H}_{1P}$ are linear
superpositions of such localised states.
\item $\mc{H}_{1P}$ satisfies the effective localisation principle on scale  $L$.
\item $\mc{H}_{1P}$ is effectively preserved, on relevant timescales,
by the dynamics of the field theory.
\end{enumerate}

This definition is intentionally somewhat vague.
The imprecision of the third criterion mirrors the way in which
quasi-particles arise in solid-state physics
 --- often the quasi-particles spontaneously decay, so that the
one-particle subspace is not exactly preserved by the dynamics.
However, provided that the decay time is long compared to other relevant
timescales (such as the time taken by the quasi-particles to move between collisions)
then the quasi-particles will provide a useful concept with which to describe the
field theory.  As the decay time decreases there will come a point at
which this concept ceases to be useful, but it would be a mistake to try
to define this point exactly.

We have also made no attempt to be precise about the phrase `where $L$ is small': how small
is small?  In non-relativistic quantum mechanics, the answer is
`arbitrarily small': a (possibly overcomplete) basis can be constructed from states effectively 
localised in
arbitrarily small regions of configuration space. (The set of all Gaussians of an arbitrary fixed width,
for instance, will do nicely.) It will turn out,
however, that this is not possible in quantum field theory: here there
will turn out to be a minimum realizable size.  It is reasonable to
think of this as giving the `size' of a particle: a particle's size is
the size of the smallest region in which it can be localised.  

Is it justifiable to be this vague in our definitions?  A robust answer
would be `it works for quasi-particles, so why not?'  More
satisfactorily, we can recall that we are not looking for particles
which can be added to the basic ontology of our theory (which, granted,
does need precise definition); the basic ontology is and remains states
of $\mc{H}_\Sigma$, or equivalently, wave-functionals on \mc{S}.
Rather, we are just finding a good way to characterise certain states
with interesting properties.  Provided these states are picked out very
accurately, there is no need to worry if the accuracy isn't perfect: we
are simply looking for accurate, robust schemes by which we can
approximate the dynamics of the theory and explain phenomena.
(For a more extended, and somewhat more philosophical, defence of
this use of approximate concepts in physics, see
\citeNP{wallacestructure}.)

In any case, it is the existence of an $\mc{H}_{1P}$ simultaneously
satisfying (1), (2) and (3) which is in need of explanation.  A space
satisfying any given one of these clauses would not be particularly remarkable:
for instance, given any collection of localised states we could construct a space
satisfying (1) by taking their span, but then this space would not
generally be preserved under time-evolution; or we could construct a
space satisfying (3) by taking the collection of all states which are
time-evolutes of our given collection, but then generally not all such states 
would be linear superpositions of members of the original collection.
Furthermore, if our system satisfied (1) and (3) but not (2), we
would have no guarantee that the concept of localisation would work for
our particles as we need it to do in non-relativistic quantum mechanics
and in scattering theory: specifically, we would have no guarantee of
the existence of projections onto particles in a specific location.

In the next three sections, we will go about constructing states which
fit the definition of a particle given above.  Before embarking on this
task, though, we should address an obvious objection: that we know
perfectly well which states of a free QFT are the one-particle states,
so all that is left to do is verify that the definition holds for these
states.

The results of the ensuing calculation would, of course, confirm \emph{that} 
free QFTs have one-particle sectors; however, it would not really answer
the question of \emph{why} they do.  The more indirect approach used
here is intended to give some insight into this second question.

\section{Modal analysis of a free field}\label{modal}

This section is a mathematical analysis of the structure of classical
linear field theories; it is a common `building block' for the two
methods of reaching the one-particle subspace which will be developed in
sections \ref{coherent} and \ref{operator}.

For the sake of mathematical rigour, this section makes some use of
distribution theory (all such material can safely be skipped by any
reader who does not get nervous upon sighting a Dirac delta function).
The notation and terminology used is essentially that of \citeN{rudin},
especially chapters 6--7; in particular, use is made of Rudin's elegant
`multi-index' notation, in which 
\begin{itemize}
\item an index $\alpha$ stands for an ordered
$n$-tuple $(\alpha_1, \ldots ,\alpha_n)$ with $\alpha_i\in \mc{Z}^+$;
\item $ D^\alpha := \left(\pbp{}{x^1}\right)^{\alpha_1} \cdots
\left(\pbp{}{x^n}\right)^{\alpha_n};$
\item $|\alpha| := \alpha_1 + \cdots + \alpha_n.$
\end{itemize}

\subsection{Required properties of \mc{R}}\label{R}

Recall that the free-field theories we are considering have the field
equation (\ref{fieldeqn}), \ie
\[\frac{\mathrm{d}^2}{\mathrm{d}t^2}\phi(\vctr{x},t) + \mc{R}
\phi(\vctr{x},t) = 0.
\]

We begin our analysis with a technical digression onto the operator \mc{R} in this
field equation.  Specifically we will require the operator to have the
following properties:
\begin{enumerate}
\item \mc{R} is a continuous linear map from $C^\infty(\Sigma)$, the
space of real smooth functions on $\Sigma$, to itself.\footnote{`Continuous' means
`continuous with respect to the topology on $C^\infty(\Sigma)$ induced by the family
of semi-norms $p_N(f) = \sup \{|D^{\alpha}f(x)|:x \in \Sigma, |\alpha| \leq N\}$'; see 
\citeN[pp.\,34--36]{rudin}
for more on such topologies.}
\item \mc{R} can be extended to a self-adjoint operator on (a dense subspace of) the space
$\lsqr(\Sigma)$ of square-integrable complex functions on $\Sigma$. (We
shall identify \mc{R} with its self-adjoint extension).
\item \mc{R} is a local operator, in the sense that $\mc{R}f(\vctr{x})$ 
depends only on the values of $f$ in an arbitrarily small neighbourhood
of $\vctr{x}$.\footnote{Given the 
short-distance cutoff introduced in section \ref{renormalisation}
to make mathematical sense of interacting QFTs, the requirement of exact locality is
not really necessary: it is enough to require that $\mc{R}f(\vctr{x})$ depends significantly on
the values of $f$ only in a neighbourhood of width $\sim L_{\mbox{cut}}$, where 
$L_{\mbox{cut}}$ is the
cutoff lengthscale; anticipating the later results of this section, 
this is to require that \mc{R} is 
$L_{\mbox{cut}}$-local.}
\item The spectrum of \mc{R} is known to be real, since it is 
self-adjoint; we shall also require it to be positive and to
be bounded below by a strictly positive eigenvalue.  (In other words,
zero is not an eigenvalue of \mc{R}; hence, \mc{R} is invertible).  
\end{enumerate}

If the spectrum of \mc{R} is discrete, \mc{R} must have a complete set of
eigenfunctions, orthonormal in the $\lsqr$ inner-product
\be \langle \phi,\psi\rangle \equiv \int_\Sigma\dr{^3 \vctr{x}}
\phi^*(\vctr{x})\psi(\vctr{x});\ee
we will denote a given such set as $f_k(\vctr{x})$. (Note that since \mc{R} is 
both real and self-adjoint we can always choose
its eigenfunctions to be all real, though we shall not always do so.) The 
eigenvalue of
$f_k$ is denoted $\omega_k^2$, with $\omega_k>0$.  

We will, in fact, take a somewhat schizophrenic attitude towards the
discreteness (or otherwise) of the spectrum of \mc{R}: for conceptual
analysis it will usually be convenient to take it as discrete, but in
practical applications we will often want to take \mc{R} to be a
differential operator on $\mathrm{R}^3$, in which case the spectrum is
necessarily continuous.  We shall therefore take the usual (if somewhat 
non-rigorous)
physicist's step of assuming that moving from a discrete to a continuous
spectrum is a purely technical matter involving no change in the
conceptual situation.

Now, let $\vctr{x}$ be any point on $\Sigma$; then we can define a linear functional
$\mc{R}_\vctr{x}$ on $C^\infty(\Sigma)$ by 
$\mc{R}_\vctr{x}\cdot f=(\mc{R}f)(\vctr{x})$; the continuity of $\mc{R}$
means that $\mc{R}_\vctr{x}$ is continuous, hence is a distribution
(generalised function) over $C^\infty(\Sigma)$.   The following results
are easy consequences of distribution theory and of the locality of
$\mc{R}$:
\begin{enumerate}
\item Because \mc{R} is local, each $\mc{R}_\vctr{x}$ has support
$\{\vctr{x}\}$.
\item From theorem 6.25 of \cite{rudin} we can deduce that (in a local
chart at \vctr{x}), we can find constants $c_\alpha$ and $N$  such that $\mc{R}_\vctr{x}=
\sum_{|\alpha|\leq N}c_\alpha D^\alpha \delta_\vctr{x}$, where
$\delta_\vctr{x}$ is a Dirac delta at \vctr{x}.
\item From the continuity of \mc{R}, it follows that, in any local
chart, we can find \emph{functions} $c_\alpha(\vctr{x})$ such that, for
any $\vctr{x}$ in the chart, $\mc{R}_\vctr{x}=
\sum_{|\alpha|\leq N}c_\alpha(\vctr{x}) D^\alpha \delta_\vctr{x}.$
\item From this, we deduce the (fairly obvious) fact that $\mc{R}$ is a
differential operator.  
\end{enumerate}

If we follow the usual fiction of treating distributions as functions,
we can (formally) define a function $\mc{R}(\vctr{x},\vctr{y})$ by
\be\int_\mc{S}\dr{^3 \vctr{y}}
\mc{R}(\vctr{x},\vctr{y})f(\vctr{y})\equiv \mc{R}_\vctr{x} \cdot f;\ee
hence 
\be(\mc{R}f)(\vctr{x})=\int_\mc{S}\dr{^3 \vctr{y}}
\mc{R}(\vctr{x},\vctr{y})f(\vctr{y}).\ee
Again formally, we can think of this function as giving the matrix
elements of \mc{R} in a position basis, provided we remember that these
elements are derivatives of delta functions.

It follows from the spectral theorem that 
\be \label{spect}\mc{R}(\vctr{x},\vctr{y})= \sum_k \omega^2_k f_k(\vctr{x})
f_k^*(\vctr{y}),\ee
and that the kernels $\mc{R}^\lambda(\vctr{x},\vctr{y})$ of the
operators $\mc{R}^\lambda$ are given by
\be \mc{R}^\lambda(\vctr{x},\vctr{y})= \sum_k \omega^{2\lambda}_k f_k(\vctr{x})
f_k^*(\vctr{y}).\ee
(Again, these kernels may well be delta-functions or other such
distributions; they are not necessarily well-behaved functions.)

Fractional powers of \mc{R} will become important later in the paper, and in general 
such operators will not be exactly local even if \mc{R} is 
(the operator $\sqrt{m^2-\nabla^2}$, for instance, is known \cite{segallocal} 
to be anti-local, in the sense that for any function $f$, $\mbox{supp} 
\sqrt{m^2-\nabla^2}f \cup \mbox{supp} f$ is all of space except possibly for a 
set of points of measure zero) but they may be `approximately local'.  
We define `approximately local' as follows:
\begin{quote} An operator \mc{R} is $L$-local iff its kernel 
$\mc{R}(\vctr{x},\vctr{y})$ drops off like \mbox{$\exp(-|\vctr{x}-\vctr{y}|/L)$}
as $|\vctr{x}-\vctr{y}|$ becomes large compared with $L$ .
\end{quote}
Informally, this means that while $\mc{R}f(\vctr{x})$ does not just depend on the 
values of $f$ in an arbitrarily small neighbourhood of $\vctr{x}$, it does depend 
significantly on the values of $f$ only in a neighbourhood of width
$\sim L$.  Note that there is a certain looseness in the definition (in the phrase
`large compared with'); purely mathematically, we could replace this with `as 
$|\vctr{x}-\vctr{y}| \rightarrow \infty$' but clearly it would be against
the spirit of the definition for (say) the kernel to start dropping off
only once $|\vctr{x}-\vctr{y}|\gg 10^{30} L$.

In the next section we will prove approximate locality for an important
subclass of \mc{R} operators.

\subsection{Euclidean-invariant \mc{R}}\label{euclid}

In this section we will consider an important sub-class of $\mc{R}$
operators: those which act upon $\mathrm{R}^3$ and which are 
invariant under spatial translations and rotations.  In this context we
can establish the approximate locality of the $\mc{R}^\lambda$
operators.

The reason for requiring translation invariance is that we can work in Fourier
space: any translation-invariant operator must have 
 the exponential functions $\frac{1}{(2 \pi)^{3/2}}\exp(i \vctr{k}\cdot \vctr{x})$ as its 
eigenfunctions and so from (\ref{spect}) we must have
\be \label{spect2}\mc{R}(\vctr{x},\vctr{y})=\frac{1}{(2 \pi)^{3}}
\int \dr{^3 \vctr{k}} \exp(i \vctr{k}\cdot(\vctr{x}-
\vctr{y}))\omega^2(\vctr{k}).\ee
Since $\mc{R}(\vctr{x},\vctr{y})$ is a rotationally-invariant sum of
derivatives of delta functions, it follows that the function
$\omega^2(\vctr{k})$ is a polynomial in $\vctr{k}\cdot \vctr{k}$.
Formally, then, the integral (\ref{spect2}) can be transformed to 
\be \label{spect3}\mc{R}(r)=\frac{1}{(2\pi)^2 i r}\int_{\mathrm{R}} \dr{k} k 
\,\omega^{2 \lambda}\exp(ikr)\ee
where $r=|\vctr{x}-\vctr{y}|$ and where we have replaced $\mc{R}(\vctr{x},\vctr{y})$ 
with 
$\mc{R}(r)$ to indicate that $\mc{R}$ depends on $\vctr{x}$ and $\vctr{y}$ only through $r$.  
For positive $\lambda$ at least, this integral is divergent, 
indicating that $\mc{R}(r)$ is distributional; however, if the spectrum is 
unbounded then for sufficiently negative 
$\lambda$ then the integral becomes convergent.   

Now, general powers of polynomials are holomorphic except along branch
cuts from the zeroes of the polynomial (for details, see any introductory
complex analysis textbook; \citeN{priestley} gives a clear exposition); furthermore, by assumption (4) of section \ref{R} we know that
the function $\omega^2(k)$ has no real zeroes.  We are then able to
evaluate (\ref{spect3}) by contour integration, as follows (this is a
simple generalisation of standard methods in QFT; see, \egc,
\citeN{peskinschroeder} for a presentation of the method for $\mc{R}=m^2-
\nabla^2$): we construct branch points from each zero in the positive
half-plane upwards parallel to the imaginary axis, and from each zero in
the negative half-plane downwards parallel to the imaginary axis.  We
then deform  the contour upwards towards $+i \infty$, wrapping it around
each branch cut in the upper half-plane in the process.  The integral is
then a sum of integrals along both sides of each branch cut in the upper
half-plane.

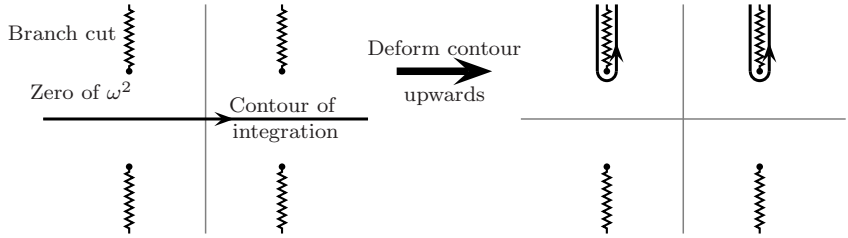
\begin{pspicture}(0,0)(4.5in,1.5in)
\psset{xunit=0.05in,yunit=0.05in}
\psset{coilarm=0.05,coilwidth=0.1}
\psline[linecolor=gray,linewidth=0.02](3,15)(37,15)
\psline[linecolor=gray,linewidth=0.02](20,3)(20,27)
\psdots[dotstyle=*](12,20)(28,20)(12,10)(28,10)
\rput(5,24){\footnotesize{Branch cut}}
\rput(7,18){\footnotesize{Zero of $\omega^2$}}
\rput(28,16.5){\footnotesize{Contour of}}
\rput(28,13.5){\footnotesize{ integration}}

\pszigzag(12,20)(12,27)
\pszigzag(28,20)(28,27)
\pszigzag(12,10)(12,3)
\pszigzag(28,10)(28,3)

\psline[linecolor=black,linewidth=0.04]{->}(3,15)(23,15)
\psline[linecolor=black,linewidth=0.04](22,15)(37,15)

\rput(45,22.5){\footnotesize{Deform contour}}
\rput(45,17.5){\footnotesize{upwards}}
\psline[linecolor=black,linewidth=0.1]{->}(40,20)(50,20)

\psline[linecolor=gray,linewidth=0.02](53,15)(87,15)
\psline[linecolor=gray,linewidth=0.02](70,3)(70,27)
\psdots[dotstyle=*](62,20)(78,20)(62,10)(78,10)

\pszigzag(62,20)(62,27)
\pszigzag(78,20)(78,27)
\pszigzag(62,10)(62,3)
\pszigzag(78,10)(78,3)

\psline[linecolor=black,linewidth=0.04](61,27)(61,20)
\psarc[linecolor=black,linewidth=0.04](62,20){0.125}{180}{360}
\psline[linecolor=black,linewidth=0.04]{->}(63,20)(63,23)
\psline[linecolor=black,linewidth=0.04](63,22)(63,27)

\psline[linecolor=black,linewidth=0.04](77,27)(77,20)
\psarc[linecolor=black,linewidth=0.04](78,20){0.125}{180}{360}
\psline[linecolor=black,linewidth=0.04]{->}(79,20)(79,23)
\psline[linecolor=black,linewidth=0.04](79,22)(79,27)

\end{pspicture}

Now, suppose that there are $N$ such branch cuts, starting at points $(u_i,v_i)$
in the upper half-plane.  Each integral along such a cut takes the form
\be I_i=\frac{\exp(-v_i r)}{(2\pi)^2ir}\int_0^{\infty}\dr{\rho} (u_i + i (v_i+\rho)) f_i(\rho) \exp(i u_i r) \exp(-
\rho r),\ee
where $f_i(\rho) = \lim_{\epsilon\rightarrow
o}(\omega^{2\lambda}(u_i+\epsilon,v_i+\rho)-\omega^{2\lambda}(u_i-
\epsilon,v_i+\rho))$. As $r \rightarrow \infty$ (or, more specifically,
as $r$ grows large compared to each $1/u_i$) the sum of all such terms will
be dominated by $I_{i_0}$, where $v_{i0}$ is the smallest of the $v_i$;
in turn, for large $r$ $I_{i_0}$ will become proportional to $\exp(-
v_{i_0}r)/r$.

This establishes that for sufficiently negative $\lambda$,
$\mc{R}^{\lambda}$ is a $1/v_{i_0}$-local operator in the sense of section
\ref{R}.  Since \emph{any} $\mc{R}^{\lambda}$ can be written as 
$\mc{R}^{\lambda-n}\mc{R}^n$ for arbitrarily large $n$, and since each
$\mc{R}^n$ is strictly local, it follows that all powers of \mc{R} are
(at least) $1/v_{i_0}$-local.  

In the case of Klein-Gordon theory, where $\mc{R}=m^2-\nabla^2$, the
only zero of the spectrum in the upper half plane is at $(0,m)$, hence
powers of $\mc{R}$ are $1/m$-local in this case.  $1/m$ is generally
referred to as the \textit{Compton wavelength}; we will extend this term
to all translation-invariant \mc{R}, and define the Compton wavelength
$L_c$ of such an \mc{R} as equal to $v_{i_0}$.

Given these results, in addition to axioms (1-4) of section \ref{R}, we will require \mc{R} 
to satisfy one of the following:
\begin{quote}5a. For all $\lambda$, \mc{R} is $L_c$-local for some $L_c$; or\end{quote}

\begin{quote}5b. \mc{R} is rotationally and translationally invariant.
\end{quote}
Of course, $5b$ implies $5a$.  

It might appear that solid-state systems do not satisfy 5b since the lattice 
structure violates translational and rotational invariance, but in fact the lattice only enters 
the observable results of the theory by imposing a short-distance cutoff, and 
hence (provided we work at lengthscales large compared with the cutoff) most 
solid-state systems may be treated as satisfying 5b.

\subsection{Modes of the free field}

Recall how to solve the free-field equation (\ref{fieldeqn}) by separation of variables: we try an ansatz of form 
$\psi(\vctr{x},t)= A(\vctr{x})B(t)$; this gives
\be A(\vctr{x}) \ddot{B}(t) + B(t) \mc{R}A(\vctr{x}) = 0.\ee
Dividing through by $A(\vctr{x})B(t)$ splits the equation into two
terms, one independent of $\vctr{x}$ and the other of $t$; this means
that the equation can be solved only by finding solutions to the paired
equations
\be\ddot{B} + \omega^2 B = 0;\ee
\be\mc{R}A = \omega^2 A\ee
where $\omega$ is to be determined. 
The second of these is simply the eigenfunction equation for $\mc{R}$. Each mode 
will have either exponential
decay/growth (for $\omega^2<0$), or sinusoidal variation (for $\omega^2>0$), in time;  
our restriction to positive \mc{R} eliminates the former case (this is the reason for this
restriction) and we are left with a set of solutions of the form
\be \phi (\vctr{x},t) = f_k(\vctr{x}) \cos (\omega_k t)\ee and
\be \phi (\vctr{x},t) = f_k(\vctr{x}) \sin (\omega_k t).\ee  
(For the Klein-Gordon equation, the $f_\vctr{k}$ are just proportional to sine and cosine
functions $\sin(\vctr{k}\cdot \vctr{x}), \cos(\vctr{k}\cdot \vctr{x})$,
with the possible values of $\vctr{k}$ constrained by the boundary
conditions and with $\omega_k^2 = m^2+\vctr{k}\cdot \vctr{k}$.)

An arbitrary solution of the equations can be expressed as a sum of solutions of this form:
\be \label{expand}\phi (\vctr{x},t) = \sum_k\frac{1}{\sqrt{\omega_k}} \left(  q_k f_k(\vctr{x}) 
\cos (\omega_k t) +  p_k f_k(\vctr{x}) \sin (\omega_k t)\right),\ee
so that a solution is given by the collection of real numbers $(q_k$, $p_k)$.

Since the space of solutions to the field equations is in one-to-one
correspondence with the phase-space \mc{P} (via $\phi(\vctr{x})\equiv
\phi(\vctr{x},0), \pi(\vctr{x}) \equiv \dot{\phi}(\vctr{x},0))$ we can regard $(q_k,p_k)$ as
coordinatizing \mc{P}: to be specific, we have
\be \label{expandrealphi}\phi(\vctr{x}) = \sum_k\frac{1}{\sqrt{\omega_k}} q_k
f_k(\vctr{x})\ee
and
\be \label{expandrealpi}\pi(\vctr{x}) = \sum_k\sqrt{\omega_k} p_k
f_k(\vctr{x}).\ee

In fact, the choice of $\sqrt\omega_k$ factors in 
(\ref{expand}) means that they are canonical coordinates, in the sense that they obey
the Poisson-bracket relations $\pb{q_k}{q_{k'}}= \pb{p_k}{p_{k'}}=0;
\pb{q_k}{p_{k'}}= \delta_{k,k'}$ (the proofs are straightforward and
make use of the orthonormality of the $f_k$).  In these coordinates the
Hamiltonian (\ref{hamilt}) becomes
\be \label{hamilt2}\ham = \frac{1}{2} \sum_k \omega_k(p_k^2+q_k^2).\ee
Thus, subject to our restrictions on \mc{R} at the start of section \ref{R}, any free-field theory can (as is of course well-known) be expressed
as a sum of independent harmonic oscillators.  

If we define $\alpha_k = \nrm(q_k + i p_k)$ we can rewrite
(\ref{expand}) in the alternative form
\be \label{cexpand}
\phi(\vctr{x},t)= \sum_k \frac{1}{\sqrt{2\omega_k}}\left( \alpha_k f_k(\vctr{x}) 
\exp(i \omega_k t) + \alpha^*_k f^*_k (\vctr{x}) 
\exp(-i \omega_k t)\right).\ee
In fact, there is no real reason to restrict to real-valued
eigenfunctions $f_k$: the expansion (\ref{cexpand}) is just as valid for
complex eigenfunctions.  Inverting it gives
\be \label{alpha}\alpha_k = \nrm\int_\mc{S} \dr{^3\vctr{x}} f^*_k(\vctr{x}) \left(
\sqrt{\omega_k} \phi(\vctr{x}) - i
\frac{1}{\sqrt{\omega_k}}\pi(\vctr{x})\right),\ee which (following the
definition $\alpha_k = \nrm(q_k + i p_k)$ in the real-$f_k$ case)
suggests taking
\be q_k = \int_\mc{S} \dr{^3\vctr{x}}\left(\sqrt{\omega_k} \phi(\vctr{x}) 
\Re f_k(\vctr{x}) + \frac{1}{\sqrt{\omega_k}}\pi(\vctr{x}) \Im
f_k(\vctr{x})\right)\ee and
\be p_k = \int_\mc{S} \dr{^3\vctr{x}}\left( -\sqrt{\omega_k}\phi(\vctr{x}) 
\Im f_k(\vctr{x}) + \frac{1}{\sqrt{\omega_k}}\pi(\vctr{x}) \Re
f_k(\vctr{x})\right)\ee
in the complex case.
We can readily show that these are still canonical coordinates; the forms of 
(\ref{expand}--\ref{expandrealpi}) become slightly more complicated but
$\ham$ still has the form (\ref{hamilt2}).  This generalisation is useful
in the Klein-Gordon equation, for instance, as it allows us to take
$f_\vctr{k} \propto \exp(i \vctr{k}\cdot\vctr{x})$, which is usually
mathematically more convenient than working with sine and cosine
functions.

The coordinate functions $q_k$ and $p_k$ have very simple time-dependence: 
from the Poisson brackets, we have $\dot{q_k} = \omega_k p_k$ and
$\dot{p_k} = -\omega_k q_k$.  Hence the time-evolution of the system is
\be q_k(t) = q_k(0) \cos (\omega_k t) + p_k(0) \sin (\omega_k t);\ee
\be p_k(t) = p_k(0) \cos (\omega_k t) - q_k(0) \sin (\omega_k t),\ee
or equivalently
\be \alpha_k(t)= \alpha_k(0) \exp(-i \omega_k t).\ee

\section{Particles through coherent states}\label{coherent}

The first method which we will use to construct particle states begins
by constructing quantum-mechanical approximations to classically
localised field states.  In fact, it will turn out that these
approximations are not particles, but they provide a natural stepping
stone towards particles.

\subsection{Harmonic oscillator coherent states}\label{harmoniccoherent}

For a linear field theory, note that even in the classical case there are states which 
are localized to greater or lesser degrees.  For instance, plane waves are (improper) 
classical states which are not localized at all, whereas we can construct fairly localized 
wave-packets.  This is different from the case of particle quantum mechanics, where 
the classical states are perfectly localized and any loss of localization occurs only at the 
quantum level.

One way to construct a localized quantum state might be as follows: we begin by choosing 
a point in the classical phase space which is fairly localized (\eg a fairly compact classical 
wave-packet) and then try to construct a quantum state which is concentrated around this 
point.  Obviously, in the general case there is no unique way of approximating a phase-space 
point in quantum mechanics since precise localization in phase-space is not a 
well-defined quantum concept.
In the case of a harmonic oscillator, however, there is a simple prescription(\citeN{glauber}; see 
\citeN{peres} for a discussion) which generates approximations to phase-space points.  
If $\opad{a}$ is 
the creation operator for such an oscillator and \ket{0} is its ground state, then the state
\be\ket{\alpha} = \exp(-|\alpha|^2/2)\exp(\alpha \opad{a})\ket{0}\ee
has the following properties:
\begin{enumerate}
\item It is a Gaussian in both position and momentum space;
\item It is centred around $q= \sqrt{2}\,\real\alpha$ in position space, and $p = 
 \sqrt{2}\,\imag\alpha$ in momentum space;
\item In both position and momentum space, the wave packet keeps its shape under time evolution
(\ie remains a Gaussian of the same width);
\item As time passes, the centres of the Gaussian in position and momentum space evolve as 
would the position and momentum of a classical harmonic oscillator with the same Hamiltonian.
\end{enumerate}
Such a state is called a \emph{coherent state}.  Note that because of
the one-to-one correspondence between phase space and the set of
solutions to the dynamical equations, and because the coherent states
track the classical evolution of phase points, we may equally well
regard coherent states as quantum approximations to classical
\emph{solutions}.

\subsection{Field coherent states}\label{fieldcoherent}

Since the free field is (mathematically speaking) a collection of independent harmonic 
oscillators, these coherent states are an appropriate tool to construct quasi-classical 
states.  The $k$th mode has creation operator $\opad{a}_k (=\nrm(q_k - i p_k))$, and 
hence a basis for the field Hilbert space $\mc{H}_\Sigma$ is given by the states created by 
successive actions 
of the different $\opad{a}_k$ on the vacuum.  

A state localized around the $k$th mode would be
\be \op{D}_k(\alpha)\ket{\Omega} :=\exp(-|\alpha|^2/2)\exp(\alpha \opad{a}_k)\ket{\Omega} 
\ee
where the classical mode being approximated is $(\Re \,\alpha
f_k/\sqrt{\omega_k},\Im \,\alpha f_k \sqrt{\omega_k})$.
Similarly, a classical state made from a superposition of modes may be quantum-mechanically 
approximated by successive application of $\op{D}_k$ operators to the vacuum, and   
the evolution of the classical state will be tracked by the quantum wave-packet.

It is vital to keep in mind the differences between classical and quantum concepts in 
what we are doing.  Remember, we are constructing a quantum wave-packet, a complex 
functional on the space of field configurations, which is concentrated around a given 
point in configuration space.  That point itself describes a classical wave-packet, 
that is, a real function on physical, three-dimensional space.

As time passes, the quantum wave-packet will not spread out across configuration space 
but will move around keeping its shape.  Its centre will move through the configuration 
space according to the classical equations of motion, which will entail the spreading out 
through physical space of the \emph{classical} wave-packet.  Thus a coherent state will 
become less localized in physical space with time even though the quantum wave-functional keeps 
its shape (and, in particular, its width) in configuration space.

Now suppose the classical solution which we want to approximate is
\be
\phi(\vctr{x},t)= \sum_k \frac{1}{\sqrt{2\omega_k}}\left( \alpha_k f_k(\vctr{x}) 
\exp(i \omega_k t) + \alpha^*_k f^*_k (\vctr{x}) 
\exp(-i \omega_k t)\right),\ee
and that it corresponds to the phase-space point $(\phi,\pi)$;
then the corresponding coherent state is
\be \ket{\mc{C}(\phi,\pi)} =\prod_k \op{D}_k(\alpha_k)\ket{\Omega}\ee
(the order in which the $\op{D}_k(\alpha_k)$ are applied is irrelevant
as they all commute).
Writing this out explicitly, we get
\be  \ket{\mc{C}(\phi,\pi)}= \prod_k
\left(
 \exp(-|\alpha_k|^2/2) \exp(-\alpha_k \opad{a}_k) \right)
\ket{\Omega}\ee
which may equally well be written as 
\be \ket{\mc{C}(\phi,\pi)} = \exp \left( -\frac{1}{2}\sum_k|\alpha_k|^2 \right)
\exp \left( -\sum_k \alpha_k \opad{a}_k \right)
\ket{\Omega}.\ee
All we have used here is 
the elementary fact --- applicable also to commuting operators --- 
that a product of exponentials is equal to the exponential of the sum of their arguments. 

Now if we take $(\phi,\pi)$ to be an element of \mc{P} satisfying $\sum_k|\alpha_k|^2=1$, 
we can define 
\be\label{phicreate}\opad{a}_{(\phi,\pi)} = \sum_k \alpha_k \opad{a}_k\ee
and
\be\label{phicreate2}\op{D}_{(\phi,\pi)}(z) = \exp(-|z|^2/2)\exp(z \opad{a}_{(\phi,\pi)}).\ee
then we will have
\be\ket{\mc{C}(\phi,\pi)} =
\op{D}_{(\phi,\pi)}(1)\ket{\Omega}.\ee
Acting on the vacuum with $\op{D}_{(\phi,\pi)}(z)$ for higher values of $|z|$ 
creates coherent states localized around successively larger wave-packets; 
in this way, the action of $\op{D}_{(\phi,\pi)}(z)$ on the vacuum as we vary $|z|$ 
and hold $(\phi,\pi)$ fixed will 
map out a collection of states, whose span is a subspace of $\mc{H}_\Sigma$. It is 
easy to see that this space can also be spanned by those states generated from the vacuum 
by successive action of the $\opad{a}_{(\phi,\pi)}$ 
operator.  Structurally there is a strong similarity to the subspaces created by 
the $\opad{a}_k$, although of course this new subspace is not preserved by time evolution.

\subsection{Coherent states are not particles}

Can the coherent states be regarded as quantum particles?  Absolutely not.
Although there are more and less localised non-relativistic quantum
particles, and more and less localised coherent states, the two forms of
localisation are wildly different.  If $(\phi,\pi)$ and $(\phi',\pi')$
are phase-space points localised in different regions\footnote{Or even
if just $\phi$ and $\phi'$, or just $\pi$ and $\pi'$, are localised in different regions.}
of $\Sigma$
then
$(\phi+\phi',\pi+\pi')$ is a classical state which is non-localized in the
sense that it is an extended field concentrated in two separated regions; 
if there are two spatially separated waves propagating on the
surface of a pond then the excitations of that surface are in this sense
nonlocal.  And all coherent states are approximations to classical states: a
coherent state formed around $(\phi+\phi',\pi+\pi')$ is non-localized only in
the same sense as its classical progenitor.

If $\psi$ and $\psi'$ are localised wave-packets of a quantum particle,
on the other hand, then $\psi+\psi'$ is nonlocalized in a wildly different
way.  Though we may be able to regard $\psi$ and $\psi'$ as approximately
classical (approximating classical point particles), we cannot so regard $\psi+\psi'$.  
After all, if the nonlocal nature of these  wave-packets could
be understood in the classical way then the profound foundational problems of
quantum nonlocality would never have arisen.


The coherent states offer us the possibility of constructing localised
quantum states, but they are certainly not particles --- localised or otherwise.  

\subsection{Two linear structures}

The argument why coherent states will not do as particles can be summed
up by noting that the following diagram does not commute:

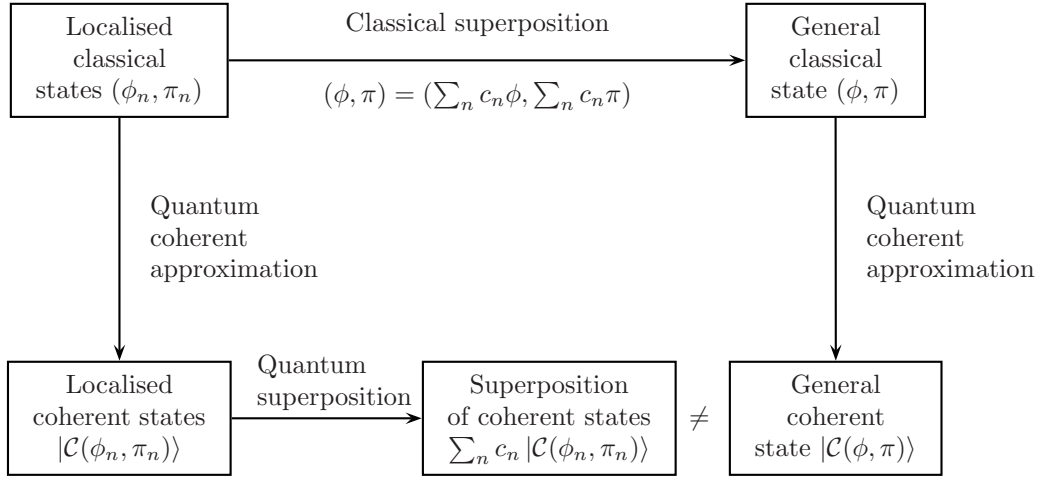
\begin{pspicture}(0,0)(4.5in,3in)
\psset{xunit=0.075in,yunit=0.0375in}

\rput(5,60){\rnode{A}
       {\rfb{Localised\\classical\\states $(\phi_n,\pi_n)$}}}
\rput(55,60){\rnode{B}
       {\rfb{ General \\ classical \\state $(\phi,\pi)$}}}

\rput(5,10) {\rnode{E}
       {\rfb{Localised \\coherent states \\ \ket{\mc{C}(\phi_n,\pi_n)}}}}
\rput(55,10) {\rnode{F}
        {\rfb{General \\coherent\\state \ket{\mc{C}(\phi,\pi)}}}}

\rput(35,10) {\rnode{G}{\rfb{Superposition \\ of coherent states \\ $\sum_n c_n 
\ket{\mc{C}(\phi_n,\pi_n)}$}}}

\rput(45.5,10){$\neq$}

\ncline{->}{A}{B}
\ncline{->}{E}{G}
\ncline{->}{A}{E}
\ncline{->}{B}{F}

\rput(30,65){Classical superposition}
\rput(30,55){$(\phi,\pi)=(\sum_n c_n \phi,\sum_n c_n \pi)$}

\rput(20,15){\begin{tabular}{l}Quantum \\superposition\end{tabular}}

\rput(13,35){\begin{tabular}{l} Quantum \\coherent \\approximation\end{tabular}}
\rput(63,35){\begin{tabular}{l} Quantum \\coherent \\approximation\end{tabular}}

\end{pspicture}

However, the strategy of approximating the classical states gives us a good method of
constructing quantum states with given locality properties, and it would
be useful to preserve that strategy in finding candidate particle states.
Clearly, what is needed is a replacement of the `quantum coherent
approximation' arrows in the diagram with some other approximation,
which still preserves localisation but which leads to a commutative
diagram. Such a replacement would have to be a far weaker concept of
approximation than that applicable to coherent states, preserving little more than localization
properties.  This is to be expected, in fact ---  non-local particle states are highly 
non-classical, so cannot be good approximations to any classical field state.

To find this replacement, consider the small-state limit of the
coherent approximation: that is, consider coherent approximations to
some state $(\lambda \phi,\lambda \pi)$, as $\lambda\rightarrow 0$.
If (without loss of generality) we take the complex coordinates $\{\alpha_k\}$ 
of $(\phi,\pi)$ to satisfy $\sum_k |\alpha_k|^2=1$, then 
from (\ref{phicreate2}), we find that
such coherent approximations are given by 
\[ \ket{\mc{C}(\lambda \phi,\lambda \pi)} =
\exp(-\lambda^2/2)\exp(\lambda\opad{a}_{(\phi,\pi)})\ket{\Omega}\]
\be\simeq \ket{\Omega} + \lambda \ket{\phi,\pi}\ee
where we have defined
\be \ket{\phi,\pi}:=\opad{a}_{(\phi,\pi)}\ket{\Omega}.\ee
Now, if the classical state $(\phi,\pi)$ is non-zero (or just non-negligible) only 
in a region \mc{A}, then 
the state $\ket{\phi,\pi}$ which we have just defined is, plausibly,
effectively localised in \mc{A}: for it is a linear combination of
\ket{\mc{C}(\phi,\pi)} (which is a coherent approximation to $(\phi,\pi)$, and
thus presumably localised in \mc{A}) and \ket{\Omega} (which, trivially,
is localised everywhere).

Of course, plausibility is at the moment all we have: as section \ref{local}
explained, effective localization is not in general a linear property.  But in
the following section we will calculate the actual localization properties
of \ket{\phi,\pi}, and find that they are indeed localized in the same region
as the classical state.  For the moment note that if such states are
localised correctly then they are precisely what we are looking for ---
for if we define $\ket{k} =
\opad{a}_k\ket{\omega}:=\ket{f_k/\sqrt{2\omega_k},0}$, we have (from
\ref{phicreate}) that
\be \ket{\phi,\pi} = \sum_k \alpha_k \ket{k}.\ee
From this it follows immediately that the \ket{\phi,\pi} states
form a subspace, and that the linear structure on that subspace mirrors
the linear structure of the classical states being approximated:
\be \opad{a}_{(A\phi+A'\phi',A\pi+A'\pi')}=A\opad{a}_{(\phi,\pi)}+A'\opad{a}_{(\phi',\pi')}.\ee

\subsection{Localization properties of \ket{\phi,\pi}
states}\label{localprops}

To investigate the localisation properties of the \ket{\phi,\pi}, we need to be able
to calculate the expectation values of products of $\op{\phi}(\vctr{x})$
and $\op{\pi}(\vctr{x})$ operators, and the major tool we shall use will
be knowledge of the (easily-calculated) commutators
\be\comm{\op{\phi}(\vctr{x})}{\opad{a}_{\phi,\pi}}= \frac{1}{2}\left(\phi(\vctr{x})+
i(\mc{R}^{-1/2}\pi)(\vctr{x})\right)\ee
and
\be\comm{\op{\pi}(\vctr{x})}{\opad{a}_{\phi,\pi}}=
\frac{1}{2}\left(\pi(\vctr{x})-
i(\mc{R}^{1/2}\phi)(\vctr{x})\right).\ee
With these known, we can calculate expectation values by 
the usual method of moving the annihilation operators over to the right where they 
annihilate $\ket{\Omega}$. For instance:
\be\matel{\phi,\pi}{\op{\phi}(\vctr{x})^2}{\phi,\pi}
\equiv\matel{\Omega}{\op{a}_{\phi,\pi}\op{\phi}(\vctr{x})^2\opad{a}_{\phi,\pi}}{\Omega}
\ee
\be= \matel{\Omega}{\op{\phi}(\vctr{x})
\op{\phi}(\vctr{x})}{\Omega}+ 
2\comm{\op{\phi}(\vctr{x})}{\opad{a}_{\phi,\pi}}
\comm{\op{\phi}(\vctr{x})}{\opad{a}_{\phi,\pi}}^*.
\ee
Subtracting off the vacuum expectation value (which is in general divergent, hence depends on whatever
high-energy cutoff procedure we have chosen to use), we get
\be\matel{\phi,\pi}{\op{\phi}(\vctr{x})^2}{\phi,\pi}
- \matel{\Omega}{\op{\phi}(\vctr{x})^2}{\Omega}
= \phi(\vctr{x})^2+
(\mc{R}^{-1/2}\pi)(\vctr{x})^2.\ee
In a similar way, we can calculate 
\be \matel{\phi,\pi}{\op{\pi}(\vctr{x})^2}{\phi,\pi}
- \matel{\Omega}{\op{\pi}(\vctr{x})^2}{\Omega}
= \pi(\vctr{x})^2+
(\mc{R}^{1/2}\phi)(\vctr{x})^2,\ee
\[ \matel{\phi,\pi}{\frac{1}{2}\op{\pi}(\vctr{x})^2+\frac{1}{2}
\op{\phi}(\vctr{x})\left(\mc{R}\op{\phi}\right)\vctr{x})}{\phi,\pi}
-
\matel{\Omega}{\frac{1}{2}\op{\pi}(\vctr{x})^2+\frac{1}{2}
\op{\phi}(\vctr{x})\left(\mc{R}\op{\phi}\right)\vctr{x})}{\Omega}\]
\be=\frac{1}{2}\pi^2(\vctr{x})+ \frac{1}{2}\left(\mc{R}^{1/2}\phi\right)^2(\vctr{x})
,\ee
\etc  (The last expectation value is that of the energy density, \ie the $(0,0)$ component
of the stress-energy tensor.)  In each case --- and, it is easy to see, for any such expectation value
--- the expectation value is
some function of the classical fields $\phi,\pi$, modified by the action
of some fractional power of $\mc{R}$.  (In particular, the energy
density is equal to the classical energy density up to the action of
such operators.)  Given that, at the end of section \ref{euclid}, $\mc{R}$ was required to 
satisfy either axiom 5b (which entails that all $\mc{R}^\lambda$ are $L_c$-local for some $L_c$) 
or axiom 5a (which requires this by fiat) it follows that 
\begin{itemize}
\item If $(\phi,\pi)$ is localised in a region \mc{A} then the difference of expectation values of 
\ket{\phi,\pi} falls off like $\exp(-d/L_c)$ with distance $d$ from \mc{A}.
\item Hence, by the definition of effective localisation (in section \ref{particle}), if 
$(\phi,\pi)$ is localised in a 
region \mc{A} then \ket{\phi,\pi} is effectively localised in the same region.
\item In view of the linearity of the map $(\phi,\pi)\rightarrow\ket{\phi,\pi}$ 
which we have constructed between classical and quantum states, it follows that the space of all states 
$\ket{\phi,\pi}$ obeys ELP on scale $L_c$.
\end{itemize}

Note that (given the definition of $L_c$-localised states) any structure that the 
classical state has on scales smaller than $L_c$ is likely to be disrupted by the action of 
$\mc{R}^\lambda$; in particular, for a classical state $(\phi,\pi)$ localised in a region small compared 
with $L_c$, the corresponding quantum state \ket{\phi,\pi} will have little in common with the classical 
state other than being effectively indistinguishable from the vacuum at distances from the 
classical state which are large compared with $L_c$.

\subsection{Particles at last}

Let us review the process we have used to construct the \ket{\phi,\pi} states.  We have taken a 
classical wave-packet and constructed a coherent state around it.  This state turns out 
to be expressible as the coherent state generated by a single creation operator, 
and in turn the action of that creation operator on the vacuum produces
a state which 
is localised in the same region as the classical wave-packet (up to variations of
size $\sim L_c$).

It is now easy to see that the following diagram commutes:

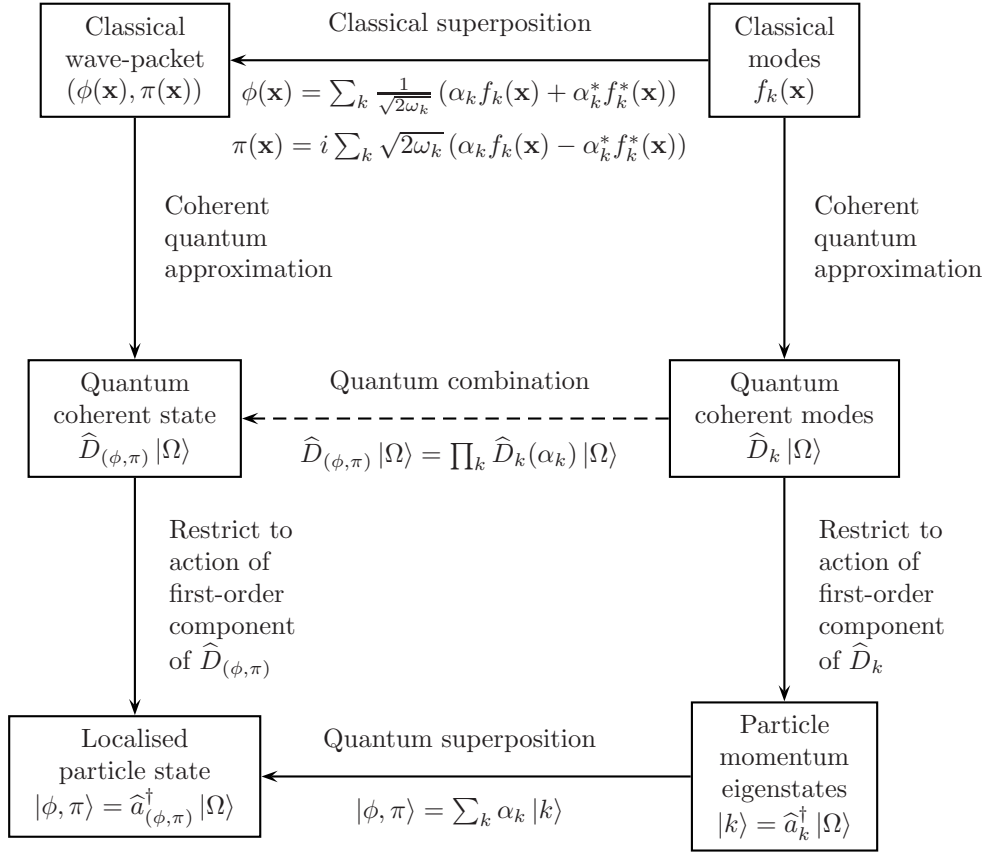
\begin{pspicture}(0.2in,0)(3.2in,4.5in)
\psset{xunit=0.085in,yunit=0.0375in}

\rput(10,110){\rnode{A}
       {\rfb{ Classical\\ wave-packet\\ $(\phi(\vctr{x}),\pi(\vctr{x})$)}}}
\rput(50,110){\rnode{B}
       {\rfb{ Classical\\ modes \\$f_k(\vctr{x})$}}}
\rput(10,60) {\rnode{C}
       {\rfb{ 	Quantum \\coherent state \\                $\op{D}_{(\phi,\pi)}\ket{\Omega}$}}}
\rput(50,60) {\rnode{D}
       {\rfb{   Quantum \\coherent modes \\
       $\op{D}_k\ket{\Omega}$}}}
\rput(10,10) {\rnode{E}
       {\rfb{Localised \\particle state \\$\ket{\phi,\pi}=\opad{a}_{(\phi,\pi)}\ket{\Omega}$}}}
\rput(50,10) {\rnode{F}
       {\rfb{Particle \\momentum \\eigenstates \\$\ket{k}=
       \opad{a}_k\ket{\Omega}$}}}

\ncline{->}{B}{A}
\ncline[linestyle=dashed]{->}{D}{C}
\ncline{->}{F}{E}
\ncline{->}{A}{C}
\ncline{->}{B}{D}
\ncline{->}{D}{F}
\ncline{->}{C}{E}

\rput(30,115){Classical superposition}
\rput(30,105){$\phi(\vctr{x})= 
\sum_k \frac{1}{\sqrt{2\omega_k}}\left(\alpha_k f_k(\vctr{x})+\alpha_k^* f_k^*(\vctr{x})\right)$}
\rput(30,98){$\pi(\vctr{x})= 
i\sum_k \sqrt{2\omega_k}\left(\alpha_k f_k(\vctr{x})-\alpha_k^* f_k^*(\vctr{x})\right)$}

\rput(30,65){Quantum combination}
\rput(30,55){$\op{D}_{(\phi,\pi)}\ket{\Omega}= \prod_k \op{D}_k(\alpha_k)\ket{\Omega}$}

\rput(30,15){Quantum superposition}
\rput(30,5){$\ket{\phi,\pi}=\sum_k \alpha_k \ket{k}$}

\rput(17,85){\begin{tabular}{l} Coherent \\ quantum \\ approximation \end{tabular}}

\rput(57,85){\begin{tabular}{l} Coherent \\ quantum \\ approximation \end{tabular}}

\rput(56,35){\begin{tabular}{l} Restrict to  \\ action of \\ first-order \\ component  \\ 
of $\op{D}_k$ \end{tabular}}

\rput(16,35){\begin{tabular}{l} Restrict to  \\ action of \\ first-order \\ component  \\ 
of $\op{D}_{(\phi,\pi)}$ \end{tabular}}

\end{pspicture}

The important properties of the diagram are:
\begin{enumerate}
\item Moving \emph{down} the diagram preserves $L_c$-localisation properties.
\item Moving from the first to the second row takes us from the
classical to the quantum regime, but does not drastically change the
nature of the states: states in the second row are good approximations
of those in the first row, in the sense spelled out in sections 
\ref{harmoniccoherent}--\ref{fieldcoherent}  This is not true for the third row: the only
sense in which \ket{\phi,\pi} approximates $(\phi,\pi)$ is that they
share the same localisation properties.  
\item Moving leftward \emph{across} the diagram corresponds to the combination of modes
to make localised states.  The middle (dashed) line is not a linear
process, but the top and bottom lines both represent linear
superposition.  However, though mathematically very similar, these
superpositions have physically very different meanings.
\item In the quantum superposition process, it is natural to consider
complex weightings for the states being superposed.  The
(mathematically) equivalent process at the classical level provides a
generalisation of the real-linear superposition process for classical
states, effectively equipping the classical solution space with a
complex structure (of which more later).
\end{enumerate}

With these results in hand, it is easy to verify that the space of
\ket{\phi,\pi} states is indeed a
one-particle space in the sense of section \ref{particle}.  The 
localised states are constructed by beginning with classically localised
wave-packets and moving down the diagram. The requirement that all
states in the space are superpositions of localised ones follows from
the equivalent property of classical phase space together with the
commutativity of the diagram. The validity of the 
superposition principle among effectively localised states is a trivial consequence of the diagram's
commutativity.  And, crucially, the closure of the one-particle subspace under 
time-evolution 
follows once we observe that the \ket{k} are energy eigenstates: this means that
the projection operator onto the one-particle subspace commutes with the Hamiltonian, so the
subspace must be preserved under time evolution.
(Equivalently, closure under time evolution follows once we observe that the map
\be (\phi,\pi) \longrightarrow \ket{\phi,\pi}\ee
commutes with time evolution.)

The essential property of the QFT which makes this whole process
possible is its linearity: without the linearity, we would not have the
\emph{classical} linear structure whose interplay with the linear
structure of $\mc{H}_\Sigma$ allowed our construction to proceed.

Henceforth, we will denote the space of all \ket{\phi,\pi} by
$\mc{H}_{1P}$.

\section{Development of the particle concept}\label{operator}

In this section we will analyse further the construction of particles
presented above.  We will examine the importance of the Compton
wavelength, and develop the links between the linear structures on phase
space and on Hilbert space; we will then use this analysis to give an
alternative way of constructing the one-particle subspace.

\subsection{Significance of the Compton wavelength}\label{compton}

The results above imply that it is localisation on the scale of the
Compton wavelength $L_c$, and not exact localisation, that is significant for
particles.  There is a straightforward physical reason for the
significance of $L_c$: as mentioned in section \ref{local}, the vacuum
state of any QFT is entangled (in the sense that field states in
different spatial regions are entangled) and this entanglement cannot be
removed from the non-vacuum states of the field without interfering with
the field's structure at energy levels comparable to the cutoff energy
(in other words, without going beyond the domain of validity of QFT).
However, the correlations in the vacuum drop off with spatial distance,
as can be seen from calculating quantities such as 
\pagebreak
\[\matel{\Omega}{\op{\phi}(\vctr{x})\op{\phi}(\vctr{y})}{\Omega}-
\matel{\Omega}{\op{\phi}(\vctr{x})}{\Omega}
\matel{\Omega}{\op{\phi}(\vctr{y})}{\Omega}\]
\be=\frac{1}{2}\mc{R}^{-1/2}\delta(\vctr{x}-\vctr{y}).\ee
If $\mc{R}^{-1/2}$ is non-local on lengthscales of $\sim L_c$, then we can
treat spatial regions separated by distances large compared with $L_c$ as
uncorrelated, but it makes rather little sense to talk about
localisation on scales small compared with $L_c$.

This also gives us at least heuristic grounds to extend the concept of
the Compton wavelength beyond Euclidean-invariant $\mc{R}$.  As was
shown in section \ref{R}, the locality of  \mc{R} requires it to have
form
\be(\mc{R}f)(\vctr{x})=\sum_{|\alpha|\leq N}
c_{\alpha}(\vctr{x})D^\alpha f(\vctr{x}),\ee
with Euclidean-invariant \mc{R} corresponding to each $c_\alpha$ being
constant.  Now, if we start with such a Euclidean-invariant \mc{R} and
introduce a very slow variation in its $c_\alpha$ (with `very slow'
meaning `significant variation on lengthscales much longer than the
Compton wavelength'), then we would expect the vacuum entanglement
lengthscale to remain substantially unchanged, which in turn suggests
that $\mc{R}^{-1/2}$ would remain non-local on the same lengthscales.

Of course, we are  using physical intuition to conjecture results of
a mathematical nature, and this conjecture is ultimately no substitute
for rigorous results about the operators $\mc{R}^{-1/2}$ in the 
non-Euclidean-invariant case.

\subsection{Field-particle duality}\label{duality}

The map 
\be (\phi,\pi) \longrightarrow \ket{\phi,\pi}\ee
defines a vector-space isomorphism\footnote{Strictly speaking the map 
takes phase space only into a proper subset of 
$\mc{H}_{1P}$, because the image of the map is not complete in the Hilbert-space
norm on by the latter (induced from $\mc{H}_\Sigma$).  Our relaxed attitude to this is due to our
approach (in section \ref{renormalisation}) to
renormalisation: we will regard $\mc{H}_{1P}$ as being cut off at some
(very high) energy, thus making it finite-dimensional and removing the
problem. \label{phasefootnote}}
between $\mc{H}_{1P}$ and the classical phase
space \mc{P}; since it commutes with time evolution, it also defines an
isomorphism between $\mc{H}_{1P}$ and the classical solution space.  
We can use this isomorphism to pull back the Hilbert-space structure
(\iec, the complex structure and the inner product) from $\mc{H}_{1P}$
to phase space, and hence to solution space: thus, we gain a
prescription by which we can regard \mc{P} as a Hilbert space.\footnote{The existence of a 
complex structure 
on the classical phase space has long been known, and in fact is a central part of 
Segal's \citeyear{segal} approach to quantisation (briefly discussed in section \ref{segal}).}

It is instructive to give the precise forms for the complex structure
(\ie, the linear operator $J$ corresponding to multiplication by $i$)
and inner product $\langle \langle \, ,  \rangle \rangle$. on \mc{P}.
We will express each in three ways:
\begin{enumerate}
\item In terms of $q_k$ and $p_k$: 
\[J(q_1,p_1; \ldots; q_k,p_k;\ldots ) = (-p_1,q_1; \ldots; -p_k,
q_k;\ldots );\]
\be \langle \langle\, (\phi,\pi),(\phi',\pi')\,\rangle \rangle
=\frac{1}{2}\sum_k[(q_k q'_k + p_k p'_k)+i(q_k p'_k - p_k q'_k)].\ee
\item In terms of $\alpha_k$:
\[J(\alpha_1, \ldots,
\alpha_k, \ldots) = (i\alpha_1, \ldots,
i\alpha_k, \ldots);\]
\be \langle \langle\, (\phi,\pi),(\phi',\pi')\,\rangle \rangle
= \sum_k \alpha_k^{*} \alpha'_k.\ee
\item Directly in terms of $(\phi,\pi)$:
\[J(\phi,\pi) = 
(-\mc{R}^{-1/2}\pi,\mc{R}^{1/2}\phi);\]
\[
\langle \langle\, (\phi,\pi),(\phi',\pi')\,\rangle \rangle
= \frac{1}{2}\int_\mc{S}\dr{^3\vctr{x}}\left(
\phi(\vctr{x})\mc{R}^{1/2}\phi'(\vctr{x}) + \pi(\vctr{x})\mc{R}^{-
1/2}\pi'(\vctr{x})\right)\]
\be
+ \frac{i}{2}\int_\mc{S}\dr{^3\vctr{x}}\left(
\pi(\vctr{x})\phi'(\vctr{x})-\phi(\vctr{x})\pi'(\vctr{x})\right),\ee
or
\be \langle \langle\, (\phi,\pi),(\phi',\pi')\,\rangle \rangle=
\langle \mc{R}^{1/4}\phi+ i \mc{R}^{-1/4}\pi,
\mc{R}^{1/4}\phi'+ i \mc{R}^{-1/4}\pi' \rangle\ee
where $\langle\,,\,\rangle$ is the ordinary $\lsqr$ inner product on
\mc{S}.
\end{enumerate}
From (2) it is easy to confirm that $J$ and $\langle\langle\, , \rangle
\rangle$ are preserved by time-evolution.  The fractional powers of
\mc{R} which occur in (3) tell us that $J$ and $\langle\langle\, , \rangle
\rangle$ are not strictly local.

Since $\mc{H}_{1P}$ and \mc{P} are isomorphic, we can use this 
Hilbert-space description of \mc{P} to provide a coordinatisation of
$\mc{H}_{1P}$ itself.  This gives us a sort of wave-function
description, albeit one in which the complex structure and inner product
are not locally defined.  Such a description is an expression, in a sense, of
wave-particle duality, with the same mathematical description applicable to
the one-particle subspace of the quantum system, and to the whole classical system.   
Note, though, that
the duality is critically dependent upon the linear structure of
the solution space --- hence we have no reason to regard field-particle
duality as a general property of field theories, but only of linear (or
nearly linear) ones.

The duality also implies that dynamics on phase space must be writeable in 
Schr\"{o}dinger-equation form.  Indeed, it is straightforward to check
that Hamilton's equations
\be \dot{\phi}=\pi; \;\; \dot{\pi}=-\mc{R}\phi \ee
are equivalent to
\be \label{pseqn}\ddt(\phi,\pi)= - J \mc{R}^{1/2} (\phi,\pi),\ee
so that the Hamiltonian is the (mildly nonlocal) operator $\mc{R}^{1/2}$.

\subsection{Alternative construction of particles: heuristic form}

We have argued that it is the interplay between two Hilbert-space structures 
--- the quantum-mechanical one and the one on the classical solution space ---
which leads to the emergence of particles, but it is perhaps somewhat
obscure exactly how that interplay comes about.  In this section and the next we will
describe an alternative way to the particle subspace which possibly
gives more insight into this question.

The basis of our new method is as follows: since the vacuum is significantly entangled
only on lengthscales of order the Compton wavelength, we would expect
that the action of a field observable like $\op{\phi}(\vctr{x})$ on the
vacuum would create a state differing from the vacuum only in the
vicinity of $\vctr{x}$ --- that is, a state $L_c$-localised at \vctr{x}.
We might further expect that, if $f$ and $g$ are real functions which vanish outside some
spatial region $\Sigma_1$, then
the state 
\be \label{oneparticle}\int \dr{^3 \vctr{x}}\left(
f(\vctr{x})\op{\phi}(\vctr{x})+g(\vctr{x})\op{\pi}(\vctr{x})\right)
\ket{\Omega}\ee
would be $L_c$-localised in $\Sigma_1$.  Furthermore, the space of all
such states (\ref{oneparticle}) is spanned by states of form
$\op{\phi}(\vctr{x})\ket{\Omega}$ and $\op{\pi}(\vctr{x})\ket{\Omega}$ 
--- that is, by states which we expect to be $L_c$-localised at a point.

It should be stressed that it is by no means obvious that these `expectations'
will be confirmed: the Reeh-Schlieder theorem reminds us that the link
between the localisation properties of operators and of those states
created by the action of those operators on the vacuum is rather
subtle.  Nonetheless, if they \emph{are} confirmed then the space of
states of the form (\ref{oneparticle}) satisfies our first two criteria (on page
\pageref{particledef}) 
for a particle subspace: that the ELP holds for the space, and that the
space is spanned by a basis of localised states.  (And in fact they can be confirmed.)

No use has yet been made of the linearity of the classical solution
space, so it is perhaps unsurprising that this property is essential to (amongst other things)
ensure that our space is to satisfy the third criterion for a one-particle space:
that the space is at least approximately preserved under time evolution.
For the linearity of the classical field equations is equivalent to the
requirement that the Hamiltonian is quadratic in the fields and their
conjugate momentum, and this in turn entails that
(writing $\op{U}(t)=\exp(-i \op{H}t)$ for the time-translation operator),
\be \op{U}(t)\op{\phi}(\vctr{x})\opad{U}(t)
\equiv \op{\phi}(\vctr{x}) - i t \comm{\op{H}}{\op{\phi}(\vctr{x})}
-t^2 \comm{\op{H}}{\comm{\op{H}}{\op{\phi}(\vctr{x})}} + \ldots\ee
is a linear combination of $\op{\phi}$ and $\op{\pi}$ operators ---
hence, the time evolution of a state like
$\op{\phi}(\vctr{x})\ket{\Omega}$ is a state of form
(\ref{oneparticle}).  This clearly entails the closure of the space of
states of form (\ref{oneparticle}) under time-evolution; hence, it is a
one-particle space.

\subsection{Alternative construction of particles: technical details}

Let us develop the formal details of this sketch.  We begin in classical
mechanics: recall that a classical observable is a real functional on
the phase space (so in one-particle mechanics the position observable
assigns to each phase-space point its configuration-space position,
etc.)  It will be important to preserve, in the following, the
distinction between the phase-space point $(\phi,\pi)$, which is a state
of the classical field, and the classical observables $\phi(\vctr{x})$ and
$\pi(\vctr{x})$, which are \emph{functionals} on the space of field
states: the action of $\phi(\vctr{x})$ on a field state returns the
field strength at the point \vctr{x}.\footnote{Similarly, in particle mechanics there
is a distinction between a particle's position \vctr{x} and the observables $\vctr{x}^i$, which
are functionals returning the $i$th component of a particle's position.}  To make this distinction 
clear, in this section we will distinguish classical observables by
writing them with a bar on top of them: $\ob{\phi(\vctr{x})}$, for
instance.  Thus, the observables $\ob{\phi(\vctr{x})}$ and $\ob{\pi(\vctr{x})}$
are defined by
\be \ob{\phi(\vctr{x})}[(\phi,\pi)]=\phi(\vctr{x})\ee
and
\be \ob{\pi(\vctr{x})}[(\phi,\pi)]=\pi(\vctr{x}).\ee

The space of observables in any classical-mechanical system has a
natural linear
structure: $(\ob{A}+\ob{B})[\vctr{v}]:=\ob{A}[\vctr{v}]+\ob{B}[\vctr{v}]$.
Furthermore, we can define time evolution for observables as follows:
if, for any phase-space point \vctr{v}, $\vctr{v}(t)$ denotes where that
phase-space point has moved to after time $t$, then we define
\be \ob{A}(t)[\vctr{v}]:=\ob{A}[\vctr{v}(t)].\ee  If \ob{H}
is the classical Hamiltonian then we can write $\ob{A}(t)$ in the
symbolic form
\be\ob{A}(t) = \exp\left(-t\pb{\ob{H}}{\cdot}\right)\ob{A},\ee
(which is to be understood as denoting the power-series expansion
\be \ob{A}(t)= \left(\sum_{n=0}^\infty \frac{(-t)^n}{n!}\pb{\ob{H}}{\cdot}\right)\ob{A}\ee
where $\pb{\ob{H}}{\cdot}\ob{A}:=\pb{\ob{H}}{\ob{A}}.$)
This movement of the dynamics from the states to the observables is
very similar, both conceptually and mathematically, to the move from the
Schr\"{o}dinger to the Heisenberg picture in quantum mechanics.  It is discussed in
more detail by \citeN[p$.20$ \textit{et seq}.\,]{woodhouse}.

Although we generally regard observables as real functionals, there is
nothing to prevent us expanding the class of observables to include
complex functionals, and we shall do so from here on. Any complex
observable, of course, has the form $\ob{A}+ i \ob{B}$, where $\ob{A}$
and \ob{B} are real observables.

So far, everything we have said about observables applies to any
classical system.  If, however (as in the case of linear field theory)
the classical phase space has a linear structure which is preserved under time
evolution, then it is possible to 
define \emph{real-linear observables}
as those real observables which are real-linear functionals; since the linear
structure commutes with time evolution on phase space then it is easy to
verify that it commutes with time evolution on the space of linear
observables as well.  In the specific case of classical fields, the
linear observables are precisely those of form
\be\int_\Sigma \dr{^3\vctr{x}}\left(
f(\vctr{x})\ob{\phi}(\vctr{x})+g(\vctr{x})\ob{\pi}(\vctr{x})\right)\ee
with $f$ and $g$ real functions.  We \label{oplocdef} will refer to any such observable
as $L_c$-localised in a region $\Sigma_1$ if the functions $f$ and $g$ are
$L_c$-localised in that region.

Since, as we have seen (in section \ref{duality}), the classical phase space can be given the
structure of a complex Hilbert space,\footnote{As was mentioned briefly before (in 
 footnote \ref{phasefootnote} on page \pageref{phasefootnote}) this
is strictly correct only if we complete the classical phase space in the norm generated by
its inner product.} we can also define 
complex, conjugate-linear observables in the obvious way (\ie all those 
complex observables which are real-linear and which satisfy
$\ob{C}[J (\phi,\pi)]= -i \ob{C}[(\phi,\pi)].$)  It is easy to
see that each such observable can be written as $\ob{C}=\ob{A}+i\ob{B}$,
with \ob{A} and \ob{B} real-linear observables.

Now, it is a basic result of Hilbert space theory (the Riesz representation theorem) 
that the space of conjugate-linear functionals on a Hilbert space is isomorphic to that
Hilbert space: in other words, that to each conjugate-linear functional
$\Lambda$ there corresponds a unique vector $\vctr{u}$ such that, for
any \vctr{v}, $\Lambda [\vctr{v}]=\langle\vctr{u},\vctr{v}\rangle$.
Hence, the space of complex conjugate-linear observables over the 
phase space of a classical linear field, conceived as a Hilbert space, 
is Hilbert-space-isomorphic to the phase space itself.  This
isomorphism takes the somewhat awkward form 
\be(\phi,\pi)\longleftrightarrow
\int_\Sigma \dr{^3\vctr{x}}\left(
(\mc{R}^{1/2}\phi(\vctr{x})-i\pi(\vctr{x}))\ob{\phi}(\vctr{x})+
(\mc{R}^{-1/2}\pi(\vctr{x})+i \phi(\vctr{x}))\ob{\pi}(\vctr{x})\right)
\ee
from which we can see that it preserves $L_c$-locality.  It takes a far
simpler form (albeit one in which its locality is obscured) if we use
the $\alpha_k$ coordinates for $(\phi,\pi)$: if we define the observable
$\ob{a}^*_k$ by 
$\ob{a}^*_k[\{\alpha_1, \ldots \alpha_n, \ldots\}]:=\alpha_k$
then the isomorphism is
\be\{\alpha_1, \ldots \alpha_n, \ldots\}
\longleftrightarrow \sum_k \alpha_k \ob{a}^*_k.\ee

The move from classical to quantum mechanics can be thought of as an
algorithm taking classical observables to quantum operators; it cannot
be applied to all observables but in field theory it does apply to all
of the real-linear observables, being generated by the maps
\be\ob{\phi}(\vctr{x}) \longrightarrow \op{\phi}(\vctr{x})\ee
and
\be\ob{\pi}(\vctr{x}) \longrightarrow \op{\pi}(\vctr{x}).\ee
The quantization map is linear, commutes with time evolution, preserves locality (trivially)
and can be extended to complex conjugate-linear observables in the obvious way:
$\ob{A}+i\ob{B} \longrightarrow \op{A}+i \op{B}.$ As might be expected,
it can be written as
\be \sum_k \alpha_k \ob{a}^*_k \longrightarrow \sum_k \alpha_k
\opad{a}_k.\ee

We have now moved from classical states, to classical observables, to
quantum observables.  Finally, we move to quantum states by applying the
appropriate quantum observable to the vacuum state \ket{\Omega}.  If the vacuum is 
entangled on lengthscales of $L_c$, then we expect (though it is not a priori guaranteed) 
that this map, too, is $L_c$-local.

This sequence of three linear isomorphisms, each commuting with time
evolution, can be summarised in the following diagram.

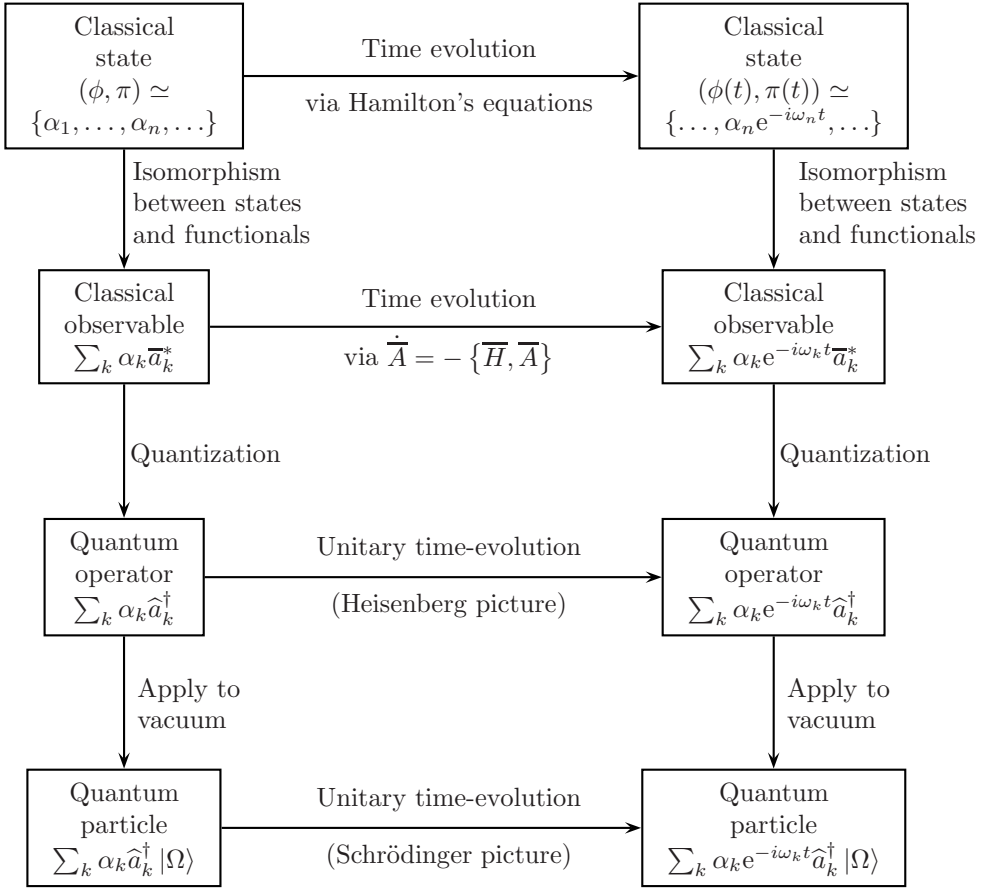
\vspace{0.2in}

\begin{pspicture}(0.45in,-0.5in)(3.45in,4.5in)
\psset{xunit=0.085in,yunit=0.0375in}

\rput(10,110){\rnode{A}
       {\rfb{ Classical\\ state
\\$(\phi,\pi)\simeq $
\\ $\{\alpha_1, \ldots, \alpha_n, \ldots\} $}}}


\rput(50,110){\rnode{B}
       {\rfb{ Classical\\ state 
\\$(\phi(t),\pi(t))\simeq$ 
\\ $\{\ldots, \alpha_n
	\mathrm{e}^{-i\omega_n t}, \ldots\} $}}}


\rput(10,75) {\rnode{C}
       {\rfb{ 	Classical \\observable \\  $\sum_k \alpha_k \ob{a}^*_k$}}}
\rput(50,75) {\rnode{D}
       {\rfb{   Classical \\observable \\ $\sum_k \alpha_k \mathrm{e}^{-i\omega_k t}\ob{a}^*_k$}}}

\rput(10,40) {\rnode{E}
       {\rfb{Quantum \\operator \\$\sum_k \alpha_k \opad{a}_k$}}}
\rput(50,40) {\rnode{F}
       {\rfb{Quantum \\operator \\$\sum_k \alpha_k \mathrm{e}^{-i\omega_k t}\opad{a}_k$}}}

\rput(10,5) {\rnode{G}
       {\rfb{Quantum \\particle \\$\sum_k \alpha_k \opad{a}_k\ket{\Omega}$}}}
\rput(50,5) {\rnode{H}
       {\rfb{Quantum \\particle \\$\sum_k \alpha_k \mathrm{e}^{-i\omega_k t}\opad{a}_k\ket{\Omega}$}}}

\ncline{->}{A}{B}
\ncline{->}{C}{D}
\ncline{->}{E}{F}
\ncline{->}{G}{H}
\ncline{->}{A}{C}
\ncline{->}{B}{D}
\ncline{->}{D}{F}
\ncline{->}{C}{E}
\ncline{->}{E}{G}
\ncline{->}{F}{H}

\rput(30,114){Time evolution}
\rput(30,106){via Hamilton's equations}

\rput(30,79){Time evolution}
\rput(30,71){via $\dot{\ob{A}}= - \pb{\ob{H}}{\ob{A}}$}

\rput(30,44){Unitary time-evolution}
\rput(30,36){(Heisenberg picture)}

\rput(30,9){Unitary time-evolution}
\rput(30,1){(Schr\"{o}dinger picture)}

\rput(16,92){\begin{tabular}{l} Isomorphism \\ between states \\ and functionals \end{tabular}}

\rput(57,92){\begin{tabular}{l} Isomorphism \\ between states \\ and functionals \end{tabular}}

\rput(15,57){Quantization}
\rput(55,57){Quantization}

\rput(54,22){\begin{tabular}{l} Apply to  \\ vacuum  \end{tabular}}
\rput(14,22){\begin{tabular}{l} Apply to  \\ vacuum  \end{tabular}}

\end{pspicture}

The important properties of the diagram are:
\begin{itemize}
\item Each map on the diagram is linear; hence, the linear structure on classical phase space
transfers to the one-particle subspace.
\item Nonetheless, the linear structures mean very different things:
the linear structures on the classical and quantum observables are
conceptually closely related to one another, but are conceptually different from the
linear structure on the classical phase space, which in turn is
conceptually different from the linear structure on the one-particle
space.
\item The vertical maps preserve $L_c$-localization; hence the one particle subspace satisfies 
the effective localisation principle on scale $L_c$ and is spanned by
states which are $L_c$-localised at a point.
(This is true by definition between the first and second lines (cf.\, page \pageref{oplocdef}),
and trivial to show between the second and third lines.  The substantive step is the one between the
third and fourth lines --- between quantum operators and quantum steps.  It is made plausible
by the observations of section \ref{compton} (that the vacuum is significantly entangled only
on lengthscales of $\sim L_c$), and proved by the calculations of section \ref{localprops}.)
\item The diagram commutes; in other words, time evolution commutes with each horizontal map. 
This ensures that states within the one-particle subspace remain in that
space, \ie that particle states are taken to particle states under time
evolution.
\end{itemize}

Compared with the coherent-states method used earlier to construct
particles, this new method has the advantage of showing more clearly how
the linear structure transfers from classical to quantum states;
however, it provides much weaker reasons for believing that the map
between classical and quantum states preserves $L_c$-locality.

(Of course, even in the coherent-state method, the argument given for
particle locality was only heuristic, relying on the assumption that if
\ket{\mc{C}} is a coherent state $L_c$-localised in some region, then
$\ket{\mc{C}}-\ket{\Omega}$ is $L_c$-localised in the same region.
Ultimately, the only way to check $L_c$-locality is by direct calculation.)

\section{The Newton-Wigner representation of
$\mc{H}_{1P}$}\label{newtonwigner}

In this section, we will consider the so-called `Newton-Wigner'
definition of local states, which we will generalise from 
Lorentz-invariant QFTs to the linear QFTs discussed above.

\subsection{Mathematical argument for the Newton-Wigner representation}

Although the representation of $\mc{H}_{1P}$ given in the previous
section is a sort of configuration-space representation, it is somewhat
awkward to use compared to the configuration-space representations which
we are accustomed to in non-relativistic quantum mechanics ---
essentially because of the non-local nature of the inner product and
complex structure.   

Purely for mathematical convenience, it would be useful to find a way of
transforming our current representation into one where the inner product
and complex structure are represented as local operations --- such a
transformation would inevitably have to be nonlocal itself, but since we
realise that the particle states are unavoidably mildly nonlocal (on
scales of $L_c$) this is not problematic.
Finding such a transformation is straightforward: it is given by
\be (\phi,\pi) \rightarrow \mc{R}^{1/4}\phi - i \mc{R}^{-1/4}\pi,\ee
or, in terms of the $\alpha_k$ coefficients, by
\be\ket{\phi,\pi}=\sum_k\alpha_k\ket{k} \rightarrow \sum_k \alpha_k f_k(\vctr{x}).\ee
It is straightforward to verify  that in the new representation, the complex
structure is just multiplication by $i$ and the inner product is the
usual $L^2$-product.  Unless $(\phi,\pi)$ is localised in a region small
compared with $L_c$ then its localisation will not be significantly changed by
the transformation; hence, the new representation is an equally accurate representation
of the actual localisation properties of the quantum states.   In the
case of Klein-Gordon theory, this is known as the Newton-Wigner
representation \cite{newtonwigner} and we will adopt this nomenclature for all such
representations (\iec, including QFTs other than Klein-Gordon theory).

If we denote the transformation from the phase-space representation of
$\mc{H}_{1P}$ to the Newton-Wigner representation by \mc{N}, then its
complex-linearity is equivalent to the statement
\be i \mc{N} = \mc{N} J.\ee
Applying \mc{N} to both sides of the Schr\"{o}dinger equation (\ref{pseqn}), we get
\be \ddt \mc{N} (\phi,\pi)=- \mc{N} J \mc{R}^{1/2} (\phi,\pi)=- i \mc{R}^{1/2}\mc{N}(\phi,\pi)\ee
so that the Hamiltonian in the Newton-Wigner representation is again
$\mc{R}^{1/2}.$  In the specific case of Klein-Gordon theory, we have
\mbox{$\mc{R}^{1/2}= (m^2-\nabla^2)^{1/2}$}, and in the non-relativistic limit
this is approximately equal to $m-\nabla^2/2m$, so that we recover the
non-relativistic free-particle Hamiltonian (up to an additive constant, the rest energy of the
particle).  It might appear that we have replaced
a nonlocal operator with a local one, but for the nonrelativistic approximation to be
valid we require that the wave-function contains only a negligible contribution from eigenstates
which don't satisfy $|\vctr{k}| \ll m$ --- which in turn means that it must be localised in a region
of size $\gg L_c$. 

\subsection{Conceptual significance of the Newton-Wigner representation}

There is another way to motivate the Newton-Wigner representation, based more on
conceptual than mathematical grounds.  In Dirac's formulation of non-relativistic quantum
mechanics, the configuration space wave-function is introduced by
\be \psi(\vctr{x}) \equiv \bk{\vctr{x}}{\psi},\ee
where the states $\ket{\vctr{x}}$ are the eigenstates of the position
operator --- so that the integral of the wave-function over a small region gives the amplitude
for the particle to be in that region.  This fits well with our description of a one-particle space 
in section  \ref{particle} as a subspace spanned by well-localised
states (indeed, in the non-relativistic case such states can be
arbitrarily well localised) and it would be useful to have a similar
representation of QFT particles.  

To what extent does the Newton-Wigner representation provide this?  The
delta functions are certainly formally equivalent to position
eigenstates, being perfectly localised in configuration space and
forming an (improper) basis for the one-particle Hilbert space.  But
obviously they are not precisely localised in real space: if
$\ket{\vctr{x}_{NW}}$ is the abstract ket corresponding to a 
delta-function at \vctr{X}, then we have
\be \ket{\vctr{x}_{NW}}=\sum_k f_k^*(\vctr{x})\ket{k},\ee
and it is easy to verify that (for instance)
\be
\matel{\vctr{x}_{NW}}{\op{\phi}(\vctr{y})\op{\phi}(\vctr{y})}{\vctr{x}_{NW}}
- \matel{\Omega}{\op{\phi}(\vctr{y})\op{\phi}(\vctr{y})}{\Omega}
\ee
is formally equal to 
\be \frac{1}{2}\left[\mc{R}^{1/4}\delta(\vctr{x}-\vctr{y})\right]^2,\ee
and hence is localised only within a region of size $\sim L_c$.

This is not a problem, however.  Recall that even in non-relativistic
quantum mechanics, no particle can actually be placed in an eigenstate
of position (such states are not even in Hilbert space).  Rather, such
states are to be seen as an idealised mathematical limit of a sequence of successively
better-localised states (see, \egc, \citeN[pp.\,100--105]{cohen}).  We can apply the same approach here, with the
proviso that the successive terms of this sequence are only successively
better localised in physical space (as opposed to in Newton-Wigner
configuration space) up to a point: the point at which the terms are
localised in regions which are not large in comparison with $L_c$.  After that point
the states will remain localised in a region of size $\sim L_c$ irrespective of how 
localised the Newton-Wigner wavefunction is.
But providing that the wavefunctions which we are studying do not
themselves vary on lengthscales comparable to $L_c$, we will only need those
terms in the sequence which are large compared with $L_c$ --- hence, we may effectively use
the concept of `position eigenstate'.

Providing the Newton-Wigner wavefunction varies slowly on
lengthscales of size $L_c$, we can regard the integral of the function
over regions of this size as giving the amplitude to find it in a state
effectively localised in the region. This then allows us 
to define an effective position operator:
\be \op{\vctr{X}}_{NW} \equiv \int_\Sigma \dr{^3
\vctr{x}}\proj{\vctr{x}_{NW}}{\vctr{x}_{NW}}.\ee
This operator will perform the same task as the non-relativistic
position operator, provided that we are not interested on scales small
compared with the minimum localisation scale: its expectation value will
give the expected value of the particle's position upon measurement, and
projections built from its spectrum correspond to projections onto the
particle being in a given region (always assuming that region to be
large compared with the minimum localisation scale $L_c$).  

However, the
relationship between this position operator and the Newton-Wigner position states is 
opposite from that which holds between non-relativistic position
operators and position eigenstates (at least in some presentations).  In
the latter case, the position operator is our starting point and
position eigenstates are position eigenstates \emph{because} they are
eigenstates of the position operator.  In QFT, the Newton-Wigner
position states are approximately local because of their relationship
with the field observables, and the position operator is constructed
from them.

\subsection{``Pathological'' features of the Newton-Wigner
representation}

The Newton-Wigner
representation of Klein-Gordon particles is well known to have two
apparently pathological features: eigenstates of position do not remain
eigenstates under Lorentz transformations, and Newton-Wigner
wavefunctions spread out superluminally.  From the current perspective,
both properties are curiosities rather than pathologies.  The Newton-
Wigner representation is not covariantly defined, so there is no
mathematical reason to expect localisation to be totally unaffected by
Lorentz boosts --- but the boosts do not cause one-particle states which
are effectively localised in a given region to stop being localised in
that region.  Similarly, Newton-Wigner wave-functions \emph{do} spread
faster than light --- but the transformation between phase space \mc{P}
and the Newton-Wigner representation commutes with time evolution, so
the Newton-Wigner state continues to represent a state effectively
localised in the region occupied by a classical state which \emph{is}
propagating at subluminal speeds.   

As such, the Newton-Wigner representation is a perfectly legitimate, and
often very convenient, way of describing states in the one-particle
subspace --- but it doesn't give the exact truth of the matter as to
where particles are localised, because there isn't one: particles are
superpositions of field excitations with finite size, so any attempt to
give a wave-function description down to arbitrarily small scales is
inevitably going to be itself arbitrary at those scales.


Before ending this discussion, I should note that there is an alternative tradition 
(going back to \cite{segal}, and currently defended
by \citeN{fleming}) that takes Newton-Wigner-localised states as \emph{by definition} localised.
In this viewpoint, localisation of states is defined directly in terms of the Newton-Wigner
position operator, rather than (as here) via the localisation properties of the 
operator algebra.  Such an approach, by taking the Newton-Wigner position operator as having
fundamental significance, must of course confront the apparent pathologies of the Newton-Wigner
representation.  

This paper, however, has as a starting assumption that localisation is
defined via the operator algebra, and as such I will not discuss
Fleming's approach.  The reader interested in how (and if) the
alternative tradition can come to terms with the pathologies should see
\citeN{flemingbutterfield} as well as the recent exchange of views
between \citeN{flemingreehschlieder} and \citeN{halvorsonreehschlieder}.

\section{Comparisons with other methods}\label{comparisons}

This section is a brief comparison of the results of this paper with
two other approaches to quantum field theory: the collision theory
developed in the framework of algebraic QFT by Haag and Ruelle, and
Segal's approach to quantizing linear systems.

\subsection{Comparison with Haag-Ruelle collision theory}

Haag-Ruelle theory (discussed in \citeN{haag}, and references therein) 
is an analysis of collision theory, and of the
particle content of quantum fields, within the framework of algebraic
QFT.  Conceptually speaking this is essentially the same framework used
here (although treated with significantly more mathematical
care\footnote{See \citeN{wallaceqft} for a discussion of the relationship
between algebraic QFT and the prima facie less rigorous QFT described
here and used in mainstream physics.}): the field is taken as primary
and particles are regarded as emergent concepts; thus, this paper's
approach is complementary to, and not in conflict with, 
Haag-Ruelle theory.

A detailed description of the theory lies beyond the scope of this
paper, but essentially it applies to any Poincar\'{e}-invariant QFT
which has a discrete eigenvalue of the mass observable (other than
zero) and for which the mass spectrum of the subspace orthogonal to the vacuum has a 
lower bound away from zero (so that it implies, for instance to massive
Klein-Gordon theory).
And it establishes that at asymptotically early and late times,
\begin{enumerate}
\item The QFT behaves as a free, massive QFT;
\item Any state which lies in the $N$-particle sector of that free
theory can set off at most $N$ widely separated detectors.
\end{enumerate}

Note that the theory applies to interacting as well as to free theories;
however, it only applies at asymptotically early and late times, and
implies that any QFT is effectively free at such times; thus, it entails
a particle interpretation only for effectively free fields (which is the
case described above).  Its virtue over the approach of this paper is
its generality (it applies to fermionic as well as bosonic fields, and to
composite as well as `elementary' particles) and its ability to infer
the particle properties of the field directly from its mass spectrum,
without recourse to concepts of field-particle duality.  However, for
this very reason the close relationship  between classical solutions and
one-particle states is somewhat obscure in the Haag-Ruelle analysis.

At first sight, the generality of the Haag-Ruelle analysis might seem to
undermine the argument of section \ref{duality} that the particle
concept is essentially bound up with the linearity of the field theory
in question.  However, the analysis itself tells us that the existence
of a massive asymptotic limit of a field theory, and the existence of a
discrete nonzero mass, are really equivalent statements.  
In the context of the present paper, the Haag-Ruelle analysis should
remind us that the linear theories we are studying should be regarded as
effective, asymptotic limits of interacting field theories, and not as
fundamental fields in their own right: for instance (though technically it lies
rather beyond the scope of this paper since it deals with fermionic
particles) we would expect QCD to be asymptotically analysable in terms
of `proton' and `neutron' fields, but these bear little resemblance to
the fundamental --- and strongly interacting --- quark fields of QCD.

(Technically, the analysis of this paper might seem more general than
the Haag-Ruelle theory since the latter is confined to 
Poincar\'{e}-invariant QFTs.  However, we can extend Haag-Ruelle theory
to general translation-invariant QFTs by generalising its spectral
condition to require that the 4-momentum spectrum includes a discrete surface 
in 4-momentum space which nowhere intersects the origin (we get this for free by
Lorentz invariance in a relativistic theory, once we have specified a
discrete mass eigenstate).  It is at best unclear how to apply 
Haag-Ruelle theory to QFTs without translational symmetry; it should,
though, be acknowledged that the above analysis is also not entirely
satisfactory in such situations as we have only a heuristic argument for
the approximate locality of non-translation-invariant $\mc{R}^\lambda$ operators 
--- recall the discussion of section \ref{compton}.)

\subsection{Comparison with Segal quantization}\label{segal}

It is instructive to compare the field-quantization approach given above
with the quantization program developed by \citeN{segal} and
others (see \citeN{saunders} for a foundational discussion).  In Segal's
approach, we \emph{begin} by choosing a complex structure $J$ on the phase space
of a linear field,\footnote{
There is another approach  very closely related to Segal quantization, in which 
instead of choosing a complex structure, we complexify the phase space
and then choose a subspace of (formally) half the dimension of the
complexified phase space.  This process is at least formally equivalent
to choosing a complex structure, though there are some 
infinite-dimensional technicalities to take care of: see \citeN{waldqft} for
an exposition of this method, and especially p.\, 46 of his book for the
relationship with the Segal method.}
which
we require to have certain properties (specifically, for any states
$u,v$, we require $J$ to satisfy $\Omega(Ju,Jv)=\Omega(u,v)$, and
$\Omega(Ju,u)\geq0$, where $\Omega(\cdot,\cdot)$ is the symplectic form
on \mc{P}).  We then use $J$ to define an inner product on \mc{P},
given by
\be \langle \langle \,u,v\,\rangle\rangle \equiv \Omega(Ju,v) -i
\Omega(u,v).\ee
This (following norm-completion) makes \mc{P} into a Hilbert space (the `one-particle Hilbert
space'), on
which we can easily show that the classical dynamics are unitary.  The
Fock space generated from that Hilbert space is then taken to be the
Hilbert space of the QFT, and there is a very natural definition of the
action of the field operators upon that space.

This is effectively the reverse procedure to ours. In Segal's approach,
we begin with the Hilbert-space structure on phase-space, declare it to
be the one-particle Hilbert space, and then build the full QFT Hilbert
space $\mc{H}_\Sigma$ from that structure; in ours, we begin with
$\mc{H}_\Sigma$, identify a subspace which deserves to be called the
`one-particle' space, and then use it to give a Hilbert-space structure
to \mc{P}.

This highlights an important conceptual difference between Segal
quantization and the `canonical' method of quantization used here.  In
our approach, the method used to quantize a (bosonic) field theory is
essentially the same irrespective of whether that theory is linear:
throw the theory into Hamiltonian form and look for a map from field
observables to operators on some Hilbert space, such that Poisson
brackets become commutators.  Then if the classical theory is linear, analysis of 
the quantum theory reveals the one-particle subspace.  In Segal's
approach, on the other hand, the linearity is fundamental to the entire
quantization algorithm, and it is most unclear how nonlinear field
theories are to be accommodated.

The attitude which one may take towards this difference depends on one's
confidence in the mathematical and conceptual status of nonlinear QFTs
--- in particular, the status of the infinities which arise in the dynamics of such theories,
and of the renormalisation process used to tame them.
If (with the author; see \citeNP{wallaceqft}) one takes a relaxed attitude to 
these infinities, then the Segal process becomes an elegant curiosity: a
nice way to understand linear quantization, but of little relevance to
general QFTs.  If, on the other hand, one regards the infinities as a
pathological problem in QFT, then (as \citeN{saunders} has argued)
it may be a virtue of the Segal approach that it forces us to look for
radical ways of reformulating the quantization of nonlinear field
theories.

It is also possible to see Segal quantization as pointing to a different
view of the relationship between field and particle than the one
defended here.   As mentioned above, in Segal quantization it is necessary to 
begin by choosing a complex structure on the classical phase space, and
different choices of structure lead to one-particle spaces with
different properties.  If one believes that there is a `right' choice of
complex structure, then there is also a `right' one-particle space, and
that space is a fundamental building block when assembling the field
Hilbert space; this is at least suggestive of a more even relationship
between field and particle concepts than the field-is-primary viewpoint
advocated above.  

On the other hand, it is also reasonable to regard the choice of a
complex structure as just being a matter of taste, in which case the
importance of the one-particle subspace is downgraded.  One argument in
favour of this attitude is that in general it is very unclear what rule
could pick out the `right' complex structure: there is a unique
prescription in the case of a time-independent linear QFT (this includes
the QFTs considered above) but in, say, the presence of a time-dependent
potential, or in a curved spacetime, then there does not appear to be
any general rule by which one complex structure could be preferred over
another.  This is, of course, a variant of the last of the arguments which
(in section \ref{problems}) led us to regard field as primary in the
first place.

\section{Conclusion}\label{conclusion}

We have shown that the one-particle subspace of a free,
bosonic, massive quantum field emerges very naturally as a consequence
of that field's classical linearity, and that the interplay between
classical and quantum linear structures is such as to give those
particle states the right sort of locality properties.

It is important to remember that this sort of construction of particle
spaces is not intended to be any sort of explanation as to why the world
around us is observed to be particulate.  At best it is able to show how
such observations are compatible with the validity of QFT --- the
question of why the actual world appears to be in a particulate state is
much subtler.  It presumably requires consideration of decoherence theory
(see \citeN{anglinzurek} for some ideas along these lines), and more
generally it involves the measurement problem.

It is also important to keep in mind the limitations on our construction.  Although these results
should extend straightforwardly to multi-component fields, they do not apply to the important
cases of fermionic, and massless, fields.  In the former case we would
expect the particle concept to be more, not less, important, due to the
Pauli exclusion principle; but the fact that spacelike-separated Fermi
field operators anticommute rather than commute suggests that we can no
longer visualise the QFT Hilbert space in terms of wave-functionals on 
configuration space.  In the case of massless fields, we expect the 
particle concept to be much more subtle:\footnote{This is supported by the situation in
algebraic QFT, in which development of a satisfactory collision theory for fields without
a mass gap is at a much more rudimentary stage; see \citeN{haag} and \citeN{buchholz} for recent discussions.}
QED, for instance, possesses not only regimes in which the
electromagnetic field is analyzable in terms of free photons, but
regimes in which it should be thought of as a classical field, and
others in which it simply provides a long-range force between
nonrelativistic particles.

This brings up perhaps the most substantial limitation of this paper's approach to
particles --- and indeed of any approach which analyses particles in
terms of asymptotic behaviour.  Although it may well be the case that
there is no particulate description of high-energy phenomena except at
asymptotically early and late times, atomic and solid-state physics
--- not to mention biology and chemistry --- provide us with a wealth
of situations in which `particles' (specifically, electrons and atomic nuclei) 
are strongly interacting and yet still maintain their own particle
character.  Analysing this situation would require a deep understanding
of the nonrelativistic limit of massive and massless QFTs, and lies far
beyond this paper's scope.

Nonetheless, despite the limitations of this approach it is satisfying
to find that field-particle duality can be understood in the context of
a field ontology, and intriguing to observe the elegant and somewhat
subtle ways in which the various linear and complex structures present
intertwine to make this understanding possible.

\small
\vspace{0.5cm}

\noindent \emph{Acknowledgements}--- I would like to thank Simon Saunders, whose own work on 
the foundations of QFT prompted my interest in the subject; and Jeremy Butterfield for many 
useful comments on earlier drafts of this paper.  I have also benefitted greatly from 
conversations with Ian Aitchison, James Binney, Katherine Brading, Harvey Brown, Keith Burnett, 
Peter Morgan, and Andrew Steane.

\end{document}